\documentclass[aps,pra,twocolumn,superscriptaddress,showemail,showpacs,longbibliography]{revtex4-2}

\usepackage{amsthm}
\usepackage{amsmath}
\usepackage{latexsym}
\usepackage{amsfonts}
\usepackage{amssymb}
\usepackage{color}
\usepackage{bbm,dsfont}
\usepackage{graphicx}
\graphicspath{{./figures/}}
\usepackage{enumerate}
\usepackage{hyperref}
\usepackage{subfigure} 
\usepackage{multirow}
\usepackage{tikz}
\usetikzlibrary{quantikz}

\usepackage{placeins}
\usepackage{import}
\usepackage{enumitem}
\usepackage{xcolor}
\usepackage{physics}
\usepackage{nicefrac}

\usepackage{bbold}
\usepackage[normalem]{ulem}
\usepackage{stmaryrd}

\usepackage[T1]{fontenc}

\usepackage[british]{babel}

\usepackage[utf8]{inputenc}
\newcommand\opone{\leavevmode\hbox{\small1\kern-3.8pt\normalsize1}}

\DeclareUnicodeCharacter{00A0}{ }

\newcommand{\kb}[2]{|#1\rangle\langle#2|} 
\newcommand{\id}{\mathds{1}} 

\newcommand{\Sys}{\mathcal{S}} 
\newcommand{\Anc}{\mathcal{A}} 
\newcommand{\Env}{\mathcal{E}} 
\newcommand{\W}{\mathcal{W}}

\newcommand{\sigmax}{\operatorname{\sigma_x}}
\newcommand{\sigmay}{\operatorname{\sigma_y}}
\newcommand{\sigmaz}{\operatorname{\sigma_z}}

\renewcommand{\vec}{\mathbf}
\newcommand{\Op}[1]{\ensuremath{\mathrm{#1}}}

\begin{document}


\title{Quantum Steering on IBMQ}

\author{Lennart Maximilian Seifert} \email{lmseifert@uchicago.edu}
\affiliation{Institut f{\"u}r Theoretische Physik, Technische
  Universit{\"a}t Dresden, D-01062, Dresden, Germany}
\affiliation{Department of Computer Science, University of Chicago, Chicago, Illinois 60637, USA}

\author{Konstantin Beyer} \email{konstantin.beyer@tu-dresden.de}
\affiliation{Institut f{\"u}r Theoretische Physik, Technische
  Universit{\"a}t Dresden, D-01062, Dresden, Germany}

\author{Kimmo Luoma} \email{ktluom@utu.fi}
\affiliation{Institut f{\"u}r Theoretische Physik, Technische
  Universit{\"a}t Dresden, D-01062, Dresden, Germany}
\affiliation{Turku Center for Quantum Physics, Department of Physics and Astronomy,
University of Turku, FI-20014, Turun Yliopisto, Finland}

\author{Walter T. Strunz} \email{walter.strunz@tu-dresden.de} \affiliation{Institut
  f{\"u}r Theoretische Physik, Technische Universit{\"a}t Dresden,
  D-01062, Dresden, Germany}

\date{\today}

\begin{abstract}
We use contemporary quantum computers to experimentally investigate quantum steering of an open quantum system by measurements on its environment. On three IBMQ processors we distinguish a qubit as the open system and perform pairwise interactions with multiple environment ancillas, following a collision model approach. Different measurement strategies on the ancillas lead to different state ensembles of the open system, which are reconstructed by employing state tomography. The amount of steering within the resulting assemblages is quantified with the help of a semidefinite program. We successfully observe the presence of quantum steering in our experimental simulations, and can discriminate the different performance qualities and noise levels of the selected quantum devices.
\end{abstract}

\pacs{}

\maketitle


\section{Introduction}
Presumably the most renowned implication of quantum theory is the phenomenon of entanglement, which defies any classical explanation. In 1935, Einstein, Podolsky and Rosen proposed their famous EPR paradox pointing out the existence of non-local entangled states \cite{EinsteinCan-Quantum-Mechanical-Description-of-Physical1935}. 
In the same year, Schrödinger realized the importance of such non-classical correlations~\cite{SchrodingerDiscussion-of-Probability-Relations1935}. He coined the term steering for the phenomena of affecting the state of a particle at a distance through measurements on another particle when the two are quantum correlated  \cite{SchrodingerDiscussion-of-Probability-Relations1935}. 
While entanglement has been studied intensively since the seminal works~\cite{EinsteinCan-Quantum-Mechanical-Description-of-Physical1935,SchrodingerDiscussion-of-Probability-Relations1935}, \emph{quantum steering}, as a special kind of quantum correlations, was thoroughly defined as recently as 2007 by Wiseman et. al. \cite{wisemanSteeringEntanglementNonlocality2007, jonesEntanglementEinsteinPodolskyRosenCorrelations2007}.

Quantum steering has been studied both theoretically~\cite{uolaQuantumSteering2020,jonesEntanglementEinsteinPodolskyRosenCorrelations2007,CavalcantiExperimental-criteria-for-steering2009,SkrzypczykQuantifying-Einstein-Podolsky-Rosen-Steering2014,QuintinoInequivalence-of-entanglement-steering2015,cavalcantiQuantumSteeringReview2017,WollmannReference-frame-independent-Einstein-Podolsky-Rosen-steering2018,beyerSteeringHeatEngines2019,zhengEnhancedEntanglementAsymmetric2020,designolleGenuineHighDimensionalQuantum2021,panNonlinearSteeringCriteria2021} and experimentally~\cite{SaundersExperimental-EPR-steering-using2010,LeachQuantum-Correlations-in-Optical2010,WalbornRevealing-Hidden-Einstein-Podolsky-Rosen2011,wittmannLoopholefreeEinsteinPodolsky2012,neryDistillationQuantumSteering2020,zhaoExperimentalDemonstrationMeasurementdeviceindependent2020,wollmannExperimentalDemonstrationRobust2020}.
One specific scenario, investigated in Refs.~\cite{WisemanAre-Dynamical-Quantum-Jumps2012,BeyerCollision-model-approach-to-steering2018}, addresses an open quantum system which is steered by measurements on its environment. In~\cite{BeyerCollision-model-approach-to-steering2018} the authors used collision models to describe the dynamics of the open system, where the environment is divided into many subenvironments (\emph{ancillas}) which sequentially interact (\emph{collide}) with the open system. Both system and ancillas were modeled with qubits. They showed that local measurements on the ancillas cause the system to jump to outcome-dependent states. These possible states form distinct measurement-dependent ensembles, which can violate a steering inequality -- an equivalent to a Bell inequality that confirms the presence of quantum steering.

In the present work we want to experimentally investigate a steering scenario on contemporary quantum computers. The efforts to build machines that utilize the principle of quantum mechanics took off around the turn of the millennium, when the computational benefit for certain problems became apparent. However, performance aspects aside, these devices can also be used to simulate actual quantum mechanics, just like Feynman pictured it during his renowned lectures in 1981 \cite{FeynmanSimulating-physics-with1982}. The IBM Quantum (IBMQ) \cite{ibmq2021} research program is one of many initiatives to construct and enhance quantum computers. Some of their architectures, which are based on superconducting transmon qubits, are available for free experimental use through the IBM Quantum Experience cloud service. They have been employed as a testbed for various theoretical concepts~\cite{wang16qubitIBMUniversal2018,koppenhoferQuantumSynchronizationIBM2020,riedelgardingBellDiagonalWerner2021} as well as for experimentally relevant purposes such as the simulation of molecules~\cite{leonticaSimulatingMoleculesCloudbased2021} and the measurement of entanglement spectra~\cite{chooMeasurementEntanglementSpectrum2018}.
Furthermore, in Ref.~\cite{garcia-perezIBMExperienceVersatile2020} it has been shown that the IBMQ quantum computer is also a versatile device for the simulation of open quantum system dynamics. In this article we use IBMQ to study another important aspect of open quantum systems: the correlations between the system and its environment. 

Specifically, we implement a dephasing dynamics as a collision model and investigate the buildup of quantum correlations between a system qubit and the environmental qubits during the evolution. Dephasing dynamics is particularly interesting with respect to the system-environment correlation. It is known that any qubit dephasing process can be modeled by means of random unitaries and is therefore essentially classical~\cite{kummererEssentiallyCommutativeDilations1987,landauBirkhoffTheoremDoubly1993,buscemiInvertingQuantumDecoherence2005,perniceSystemEnvironmentCorrelations2012}. Hence, it can never be verified on the qubit system alone that a dephasing process stems from a quantum environment and is not just mimicked by a classical random process, but one needs to include the environment as well.

We use quantum steering to demonstrate the quantumness of the correlation between system and environment. In a quantum steering task one side is treated as an untrusted black box, i.e., no assumptions are made about the measurement apparatus and the dimensions of the Hilbert space (as in a Bell test). The other side remains device-dependent and therefore relies on the validity of the assumptions about the measurement device (as in an entanglement test). This asymmetry reflects the typical situation for open quantum systems: While the system of interest is supposed to be well known and characterized, the environment is generally less accessible and treating it as a black box seems appropriate. We use different ancilla measurement strategies in combination with state tomography to reconstruct the state ensemble of the open system. Quantum steering can then be quantified by a steering weight~\cite{SkrzypczykQuantifying-Einstein-Podolsky-Rosen-Steering2014,cavalcantiQuantumSteeringReview2017} calculated from the experimental data. We implement the scenario on three of the IBMQ devices and verify the existence of quantum steering.

We structure this article as follows: In Sec. \ref{sec:steering} we formally introduce the concept of quantum steering and present the quantifier of our choice; in Sec.~\ref{sec:steerin-of-open-system} we give an overview of quantum steering of an open quantum system in terms of collision models. In Sec.~\ref{sec:experiment} we describe the concrete process considered in this work. Section \ref{sec:implementation} explains the experimental implementation of the simulations with quantum circuits, which is followed by the analysis of steering from acquired data in Section \ref{sec:analysis}. Finally, we give a summary. More information on methodology and further results are provided in the appendix.

\section{Theoretical concepts}

\subsection{Quantum Steering}
\label{sec:steering}
Let us consider a general bipartite setup where two parties, Alice and Bob, share an unknown quantum state $\rho$. Alice and Bob would like to demonstrate that the shared state is quantum correlated. To this end they perform local measurements on their respective sides and communicate classically with each other. If they trust their measurement devices they can measure locally a set of informationally complete observables to reconstruct the shared state and check whether it is entangled. If they do not trust their devices, they can still verify quantum correlations in the shared state if they are able to violate a Bell inequality with the measured statistics. The former is called a device-dependent and the latter a device-independent scenario.

Quantum steering is an asymmetric semi-device-independent scenario~\cite{wisemanSteeringEntanglementNonlocality2007, jonesEntanglementEinsteinPodolskyRosenCorrelations2007}. Only Bob performs trusted quantum measurements while Alice's measurement device is treated as a black box. Alice can apply different measurement strategies labeled by $x$ on her part of the bipartite state and obtains outcome $a$ with probability $p(a|x)$ from her untrusted apparatus. By doing so she can, in general, prepare different state ensembles $\{p(a|x), \rho_{a|x}\}_x$ on Bob's side. In each run Bob randomly chooses one of Alice's strategies $x$, asks her to perform the corresponding measurement and share the outcome $a$ with him. He is then -- after many runs of the experiment -- able to reconstruct the states $\rho_{a|x}$ through state tomography. 

It is useful to introduce the corresponding subnormalized states $\sigma_{a|x}=p(a|x)\rho_{a|x}$. The set $\{\sigma_{a|x}\}$ is often called an \emph{assemblage} \cite{cavalcantiQuantumSteeringReview2017}. From now on we will drop the brackets and simply refer to $\sigma_{a|x}$ as an assemblage unless otherwise deemed necessary. We say that an assemblage allows a \emph{local hidden state model} (``$\sigma_{a|x}$ is LHS'') if it can be decomposed into the form \cite{cavalcantiQuantumSteeringReview2017}
\begin{equation}
    \label{eq:LHS_decomposition}
    \sigma^\text{LHS}_{a|x} = \int \dd{\lambda} \mu(\lambda) p(a|x,\lambda) \rho_\lambda.
\end{equation}
This implies that Bob's observations can be explained by a classical mixture of hidden states $\rho_\lambda$ on his side and the black box on Alice's side could just output this classical statistics. However, if such a decomposition cannot be found, then $\sigma_{a|x}$ demonstrates quantum steering and the shared state is said to be \emph{steerable} from Alice to Bob. Steerability implies entanglement, while on the other hand the set of steerable states is a strict superset of the Bell nonlocal states. 

An important property of general assemblages is the \emph{no-signaling condition}
\begin{equation}
    \label{eq:no-signaling}
    \sum_a \sigma_{a|x} = \sum_a \sigma_{a|x'} = \rho_B \quad \forall x, x',
\end{equation}
which expresses that averaging the ensemble states over Alice's outcomes yields the reduced state on Bob's side $\rho_B$ for all choices of measurement $x$. This is important because otherwise Alice could influence Bob's local reduced state by her measurement.

Given an assemblage $\sigma_{a|x}$, trying to find a decomposition of the form \eqref{eq:LHS_decomposition} is very difficult in general. In past works different steering inequalities have been proposed \cite{CavalcantiExperimental-criteria-for-steering2009, WisemanAre-Dynamical-Quantum-Jumps2012, WollmannReference-frame-independent-Einstein-Podolsky-Rosen-steering2018}, whose violation can show that a LHS decomposition does not exist. 

In this work we will make use of another witness of steering that is moreover able to quantify steering, the so called \emph{steering weight} SW \cite{SkrzypczykQuantifying-Einstein-Podolsky-Rosen-Steering2014, cavalcantiQuantumSteeringReview2017}. It represents the minimal amount of a generic assemblage $\gamma_{a|x}$ that needs to be added to a LHS assemblage $\sigma^\text{LHS}_{a|x}$ in order to reproduce the original assemblage $\sigma_{a|x}$:
\begin{align}
    \label{eq:SW}
    \operatorname{SW}(\sigma_{a|x}) = &\min_{\vspace{10pt} \gamma_{a|x}, \sigma^\text{LHS}_{a|x}, p} \quad p \\[2pt]
    &\text{s.t.} \quad \sigma_{a|x} = p \gamma_{a|x} + (1-p) \sigma^\text{LHS}_{a|x}. \notag
\end{align}
This optimization problem can be efficiently solved using semidefinite programming (SDP) \cite{VandenbergheSemidefinite-Programming1996, cavalcantiQuantumSteeringReview2017}. We will use the code provided by the authors of Ref.~\cite{cavalcantiQuantumSteeringReview2017} and apply this scheme for quantifying steering within our experimentally produced assemblages.

\subsection{Quantum steering of an open system}
\label{sec:steerin-of-open-system}
In the present work we analyze quantum steering in an open quantum system, i.e., a quantum system $\Sys$ coupled to an environment $\Env$. We will use steering to verify that the system becomes quantum correlated with the environment by the open system dynamics. The asymmetry of a quantum steering task very well reflects the asymmetric roles of the system and the environment. While the open system is usually considered to be well known and manageable, this generally cannot be said about the environment, whose degrees of freedom might be partially unknown. Thus, in order to verify quantum correlations between $\Sys$ and $\Env$, it is reasonable to resort to quantum steering. The open system $\Sys$ is treated as the trusted part on Bob's side and the environment $\Env$ remains untrusted so that Alice's measurements are treated as a black box~\cite{WisemanAre-Dynamical-Quantum-Jumps2012,BeyerCollision-model-approach-to-steering2018,beyerSteeringHeatEngines2019}.

We will model the open system dynamics by a collision model. Let $\rho$ denote the density operator describing the open quantum system $\Sys$. The surrounding environment is modeled by discrete subenvironments (ancillas) $\Anc_i$ initially in the state $\rho_{\Anc_i}$ which sequentially interact with the open system. Here, every such collision shall be governed by a unitary operator $\W_i$. After the collision the subenvironments do not interact again neither with the system nor with another ancilla. The joint state of the system and the environment after $N$ collisions is then given by
\begin{equation}
    \rho_{\Sys\Env} = \W_N\ldots\W_1 \qty(\rho_0 \otimes \rho_{\Anc_1} \otimes\ldots\otimes\rho_{\Anc_N}) \W_1^\dagger\ldots\W_N^\dagger,
\end{equation}
where the unitary $\W_i$ acts on the system and the $i$th ancilla only and $\rho_0$ is the initial state of the system.

To demonstrate that the system has entangled with the environment during the dynamics, Alice and Bob perform a steering task on the joint state $\rho_{\Sys\Env}$. By implementing different measurement scenarios $x$ on the ancillas, Alice tries to steer Bob's system into different ensembles that form the assemblage $\sigma_{a|x}$. After a tomography of the assemblage, Bob can calculate the steering weight $\operatorname{SW}$. For $\operatorname{SW}(\sigma_{a|x}) >0$ his system is steerable by Alice and the open system must be entangled with the environment.

Let us assume that the collision model correctly describes the quantum channel on Bob's system and that Alice is indeed able to perform quantum measurements on the ancillas. The subnormalized states in the assemblage are then theoretically given by
\begin{align}
    \sigma_{a|x} = \Tr[\rho_{\Sys\Env} (\id_\Sys \otimes A^x_a)],
\end{align}
where the $\{A^x_a\}$ form a positive operator valued measure (POVM) corresponding to measurement strategy $x$ on the environment. For practical reasons that will become clear later we restrict Alice's measurements to observables that are local on each subenvironment. Thus each POVM element is given by 
\begin{align}
    A^x_a = A^x_{a_1} \otimes \ldots \otimes A^x_{a_N},
\end{align}
where the $\{A^x_{a_i}\}$ now form a local POVM on the $i$th ancilla and the measurement outcome $a$ is a tuple of all local outcomes $a = (a_1,\ldots,a_N)$.
A measurement strategy $x$ on Alice's side can then be specified by a tuple of POVMs that defines the measurement on each ancilla, respectively.

Crucially, while Alice can use her knowledge about the joint state $\rho_{\Sys\Env}$ to tailor her measurement strategies $x$, the demonstration of steering does not depend on the validity of this information. Bob's reconstruction of the assemblage only depends on the classical labels of the measurement strategy $x$ and the outcome $a$ without any reference to the underlying state or the quantum measurements on Alice's side.

\section{The model}
\label{sec:experiment}
We will now turn to a specific realization of the setup described above that can be implemented as a qubit collision model on the IBMQ quantum computer. The interactions between the system qubit $\Sys$ and several ancillas $\Anc_i$ induce a bit flip channel --- also known as dephasing in the $\sigma_x$-basis --- on $\Sys$. Subsequently, we check if quantum correlations between the system and its environment can be experimentally verified. The bit flip channel is unital and thus could always be implemented by a random unitary process~\cite{kummererEssentiallyCommutativeDilations1987,landauBirkhoffTheoremDoubly1993,buscemiInvertingQuantumDecoherence2005,perniceSystemEnvironmentCorrelations2012}. This makes the channel interesting for our purposes because demonstrating steering shows that the channel is indeed induced by a quantum environment and not just a consequence of a classical random process on the system alone.

\subsection{Dephasing in $\sigmax$-basis}
In our collision model the ancillas shall start in the $\sigmaz$ eigenstate  $\rho_{\Anc_i}=\rho_\Anc =\kb{0}{0}$. A single collision is given by a rotation of the ancilla about the $y$-axis followed by an ancilla-controlled CNOT gate $CX^\Sys_\Anc$ targeting the system qubit:
\begin{minipage}{\linewidth}
\vspace{0.7em}
\begin{minipage}{0.35\linewidth}
\raggedright
\includegraphics[scale=0.92]{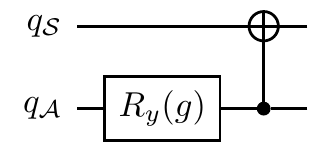}
\end{minipage} \hfill
\begin{minipage}{0.64\linewidth}
\vspace{-0.5cm}
\begin{equation}
    \longleftrightarrow \quad \W = CX^\Sys_\Anc \, \qty(\id \otimes R_y(g))
\end{equation}
\end{minipage} 
\vspace{0.7em}
\end{minipage}

The rotation angle $g \in (0,\frac{\pi}{2})$ acts as a coupling parameter and determines the strength of a single collision.
The reduced state of Bob's system qubit $\Sys$ changes during a single collision as
\begin{align}
\label{eq:single-step-map}
    \mathcal{E}(\rho_\Sys) = \Tr_\Anc[\W (\rho_\Sys\otimes\rho_\Anc) \W^\dagger].
\end{align}
For convenience we represent the state of the system qubit by its 4-component Bloch vector $\vec k$.
Using the Pauli matrices $\sigma_{1,2,3}$ and defining $\sigma_0 = \id$ we can write the system state $\rho_\Sys$ as
\begin{align}
    \rho_\Sys = \sum_{i=0}^3 k_i\, \sigma_i,
\end{align}
with $\vec{k} = (1,\vec{r})^\intercal$ and $\vec r = (\Tr[\sigma_x \rho_\Sys], \Tr[\sigma_y \rho_\Sys], \Tr[\sigma_z \rho_\Sys])^\intercal$.
The map $\mathcal{E}$ in the Bloch vector representation then reads:
\begin{equation}
    \Lambda = \mqty(1&0&0&0\\0&1&0&0\\0&0&\cos(g)&0\\0&0&0&\cos(g)).
\end{equation}
To establish a connection to a continuous dephasing process we define the time step $t = - \ln(\cos(g)) \in (0, \infty)$ and the total time after $N$ collisions as $T = N t$. Applying $N$ collision we then obtain
\begin{equation}
    \label{eq:LambdaN}
    \Lambda^N = \mqty(1&0&0&0\\0&1&0&0\\0&0&e^{-T}&0\\0&0&0&e^{-T}),
\end{equation}
which is independent of the underlying time discretization. Accordingly, our model stroboscopically simulates an exponential decay of the system's Bloch vector's $y$- and $z$-component over time, i.e., a dephasing in the $\sigmax$-basis.

Henceforth, we focus on a specific dephasing channel with fixed time $T=2.0$. As we have seen we can exactly model this channel by an arbitrary number of collisions. Physically speaking we can decide how large the environment is that contributes to the dephasing process by varying the number of ancillas involved. Due to limitations of the available IBMQ devices we will consider $1$ to $4$ collisions.

\begin{figure}
    \centering
    \includegraphics[width=\linewidth]{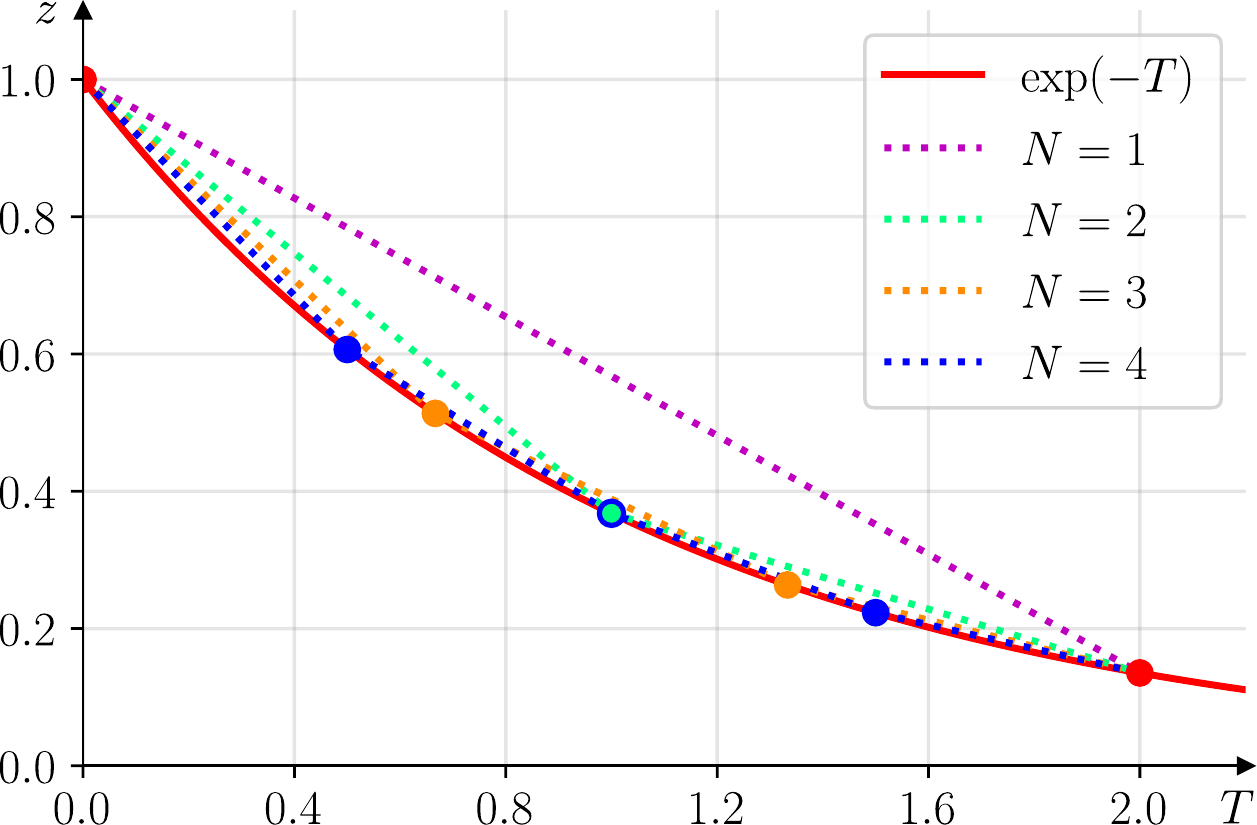}
    \caption{Exponential decay of the $z$-component of the system's Bloch vector due to dephasing in the $\sigmax$-basis. We model a $T=2.0$ process (marked by the red dots) with different numbers of discrete collisions $N$, as indicated by the dashed lines.}
    \label{fig:collisions}
\end{figure}

The system qubit shall start in the ground state $\rho_0 = \kb{0}{0}$ with Bloch vector $\vec{r}_0 = (0,0,1)^\intercal$, which is why the final reduced state predicted by quantum theory is $\vec{r} = (0,0,e^{-2})^\intercal$. The discrete evolution for each case is depicted in Figure \ref{fig:collisions} and Table \ref{tab:parameters} shows the corresponding time steps $t$ and coupling parameters $g$.

\subsection{Alice's measurement scenarios}
\label{sub:strats}
We will now establish the measurement scenarios on the ancilla qubits that we subsequently use to demonstrate quantum correlations between the system and the environment.

Producing two different state ensembles would suffice for investigating the presence of quantum steering. However, since we expect to encounter noise distorting the quantum correlations, we apply three different measurement scenarios on Alice's side in order to enhance the steering weight.

We aim to create ensembles that are as distinct as possible in the Bloch ball. Since the IBMQ devices can only measure each qubit separately we also restrict ourselves to observables that are local on each ancilla. In principle it would be possible to measure also nonlocal observables on multiple ancillas by implementing a global gate before the local measurements. However, implementing a suitable nonlocal unitary would require a large number of error-prone two-qubit gates and introduce so much noise that a steering task became infeasible.

In analogy to the notation of Section \ref{sec:steerin-of-open-system}, we may represent a measurement strategy by a sequence of POVMs $x = \qty{A_1, A_2, \dots, A_N}$. The local measurements in the IBMQ devices have binary outcomes. In the ideal case these measurements are projective and, thus, can be represented by Bloch vectors $\vec a_i$ of unit length. We assume these kind of measurements for constructing expedient scenarios. Since the local POVMs are dichotomic, it suffices to specify the elements associated with outcome ``0''. The respective element for outcome ``1'' is then given by $-\vec a_i$. Therefore, we can represent a measurement scenario on Alice's side by $x = (\vec{a}_1, \vec{a}_2, \dots, \vec{a}_N)$. We find two scenarios which lead to distinct ensembles easily:
\begin{itemize}
    \item $x_1 = \qty(\vec{a}_i = (0,0,1)^\intercal \vert i = 1, \dots, N)$
    \item $x_2 = \qty(\vec{a}_i = \qty(\sin(g),0,\cos(g))^\intercal \vert i = 1, \dots, N)$
\end{itemize}

\begin{figure*}[ht]
    \centering
    \subfigure[Strategy $x_1$]{\includegraphics[width=0.28\textwidth]{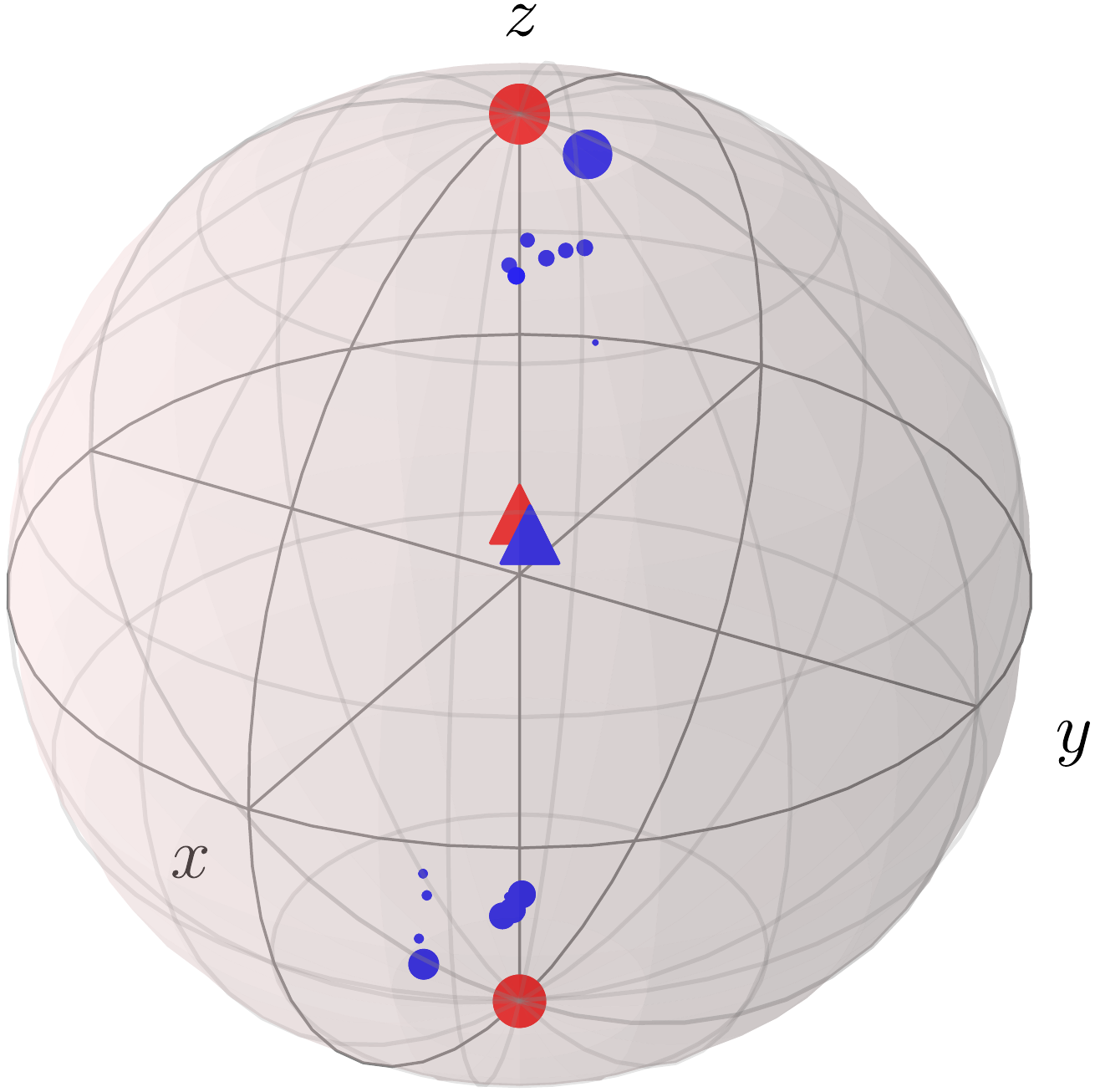} \label{fig:x1}}
    \hfill
    \subfigure[Strategy $x_2$]{\includegraphics[width=0.28\textwidth]{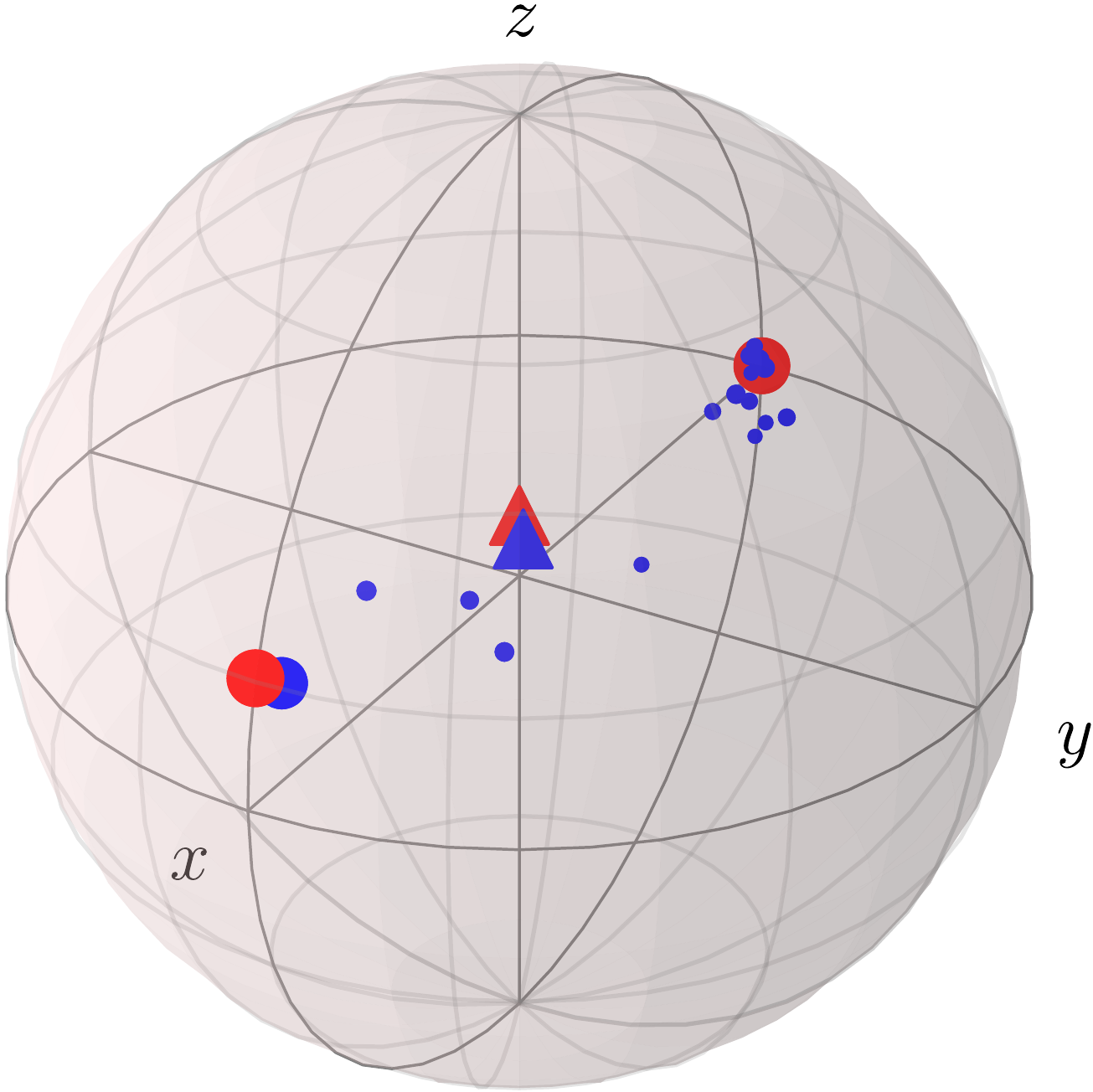} \label{fig:x2}}
    \hfill
    \subfigure[Strategy $x_3$]{\includegraphics[width=0.28\textwidth]{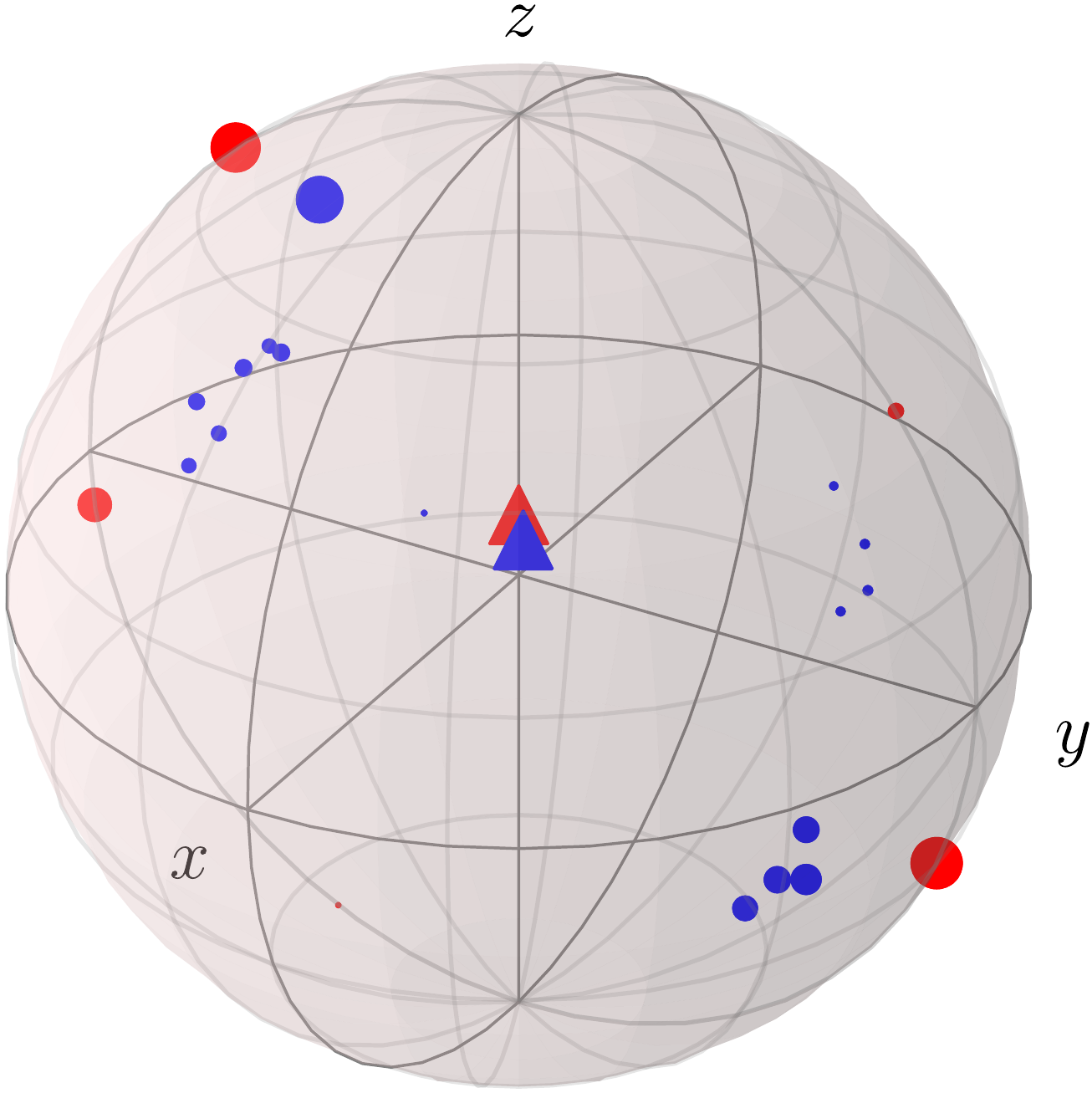} \label{fig:x3}}
    \caption{State ensembles created with Alice's measurement strategies after the $T=2.0$ dephasing process with $N=4$ collisions, juxtaposing theory (red dots) and experimental results from \texttt{ibmq\_santiago} (blue dots). For reconstructing the latter the ideal POVM set $\mathcal{B}_\text{id}$ is assumed. The different dot sizes represent the state probabilities within the ensemble, while the triangles correspond to the state averages respectively.}
    \label{fig:ensembles}
\end{figure*}

The red dots in Figures \ref{fig:x1} and \ref{fig:x2} show the so produced ensembles in Bob's system, where the dot sizes encode the state probabilities $p(a|x)$. The ensembles lie on almost orthogonal lines to each other, which makes them great candidates for demonstrating steering. Averaging over all states of an ensemble yields the reduced state due to Eq.~\eqref{eq:no-signaling}, which is represented by a red triangle.

The corresponding noiseless assemblage already maximizes the steering weight to unity theoretically.
Thus a third ensemble seems unnecessary. However, noisy assemblages show steering weights $\operatorname{SW} < 1$ and an increase may be achieved by producing more ensembles. 

For finding a suitable third strategy, we first parametrize the Bloch vectors $\vec{a}_i$ with usual spherical coordinates $(\theta_i, \varphi_i)$. Then we (theoretically) compute the full assemblage, manually add a small amount of white noise $\lambda=0.05$ (that is, apply the transformation $\vec{r} \mapsto (1-\lambda) \vec{r}$ to all states)~\footnote{Without the noise the optimization would fail since the steering weight would already be saturated by the first two pure ensembles.} and maximize the steering weight over all angles. We find that for a given number of collisions $N$, all ancillas should be measured with the same local POVM. The first angle is always $\varphi_i = \nicefrac{\pi}{2}$ while the angle $\theta_i = \theta_N$ depends on the total number of collisions $N$. The values for $\theta_N$ can be found in Table \ref{tab:parameters}. The final scenario is then given by:
\begin{itemize}
    \item $x_3 = (\vec{a}_i = \qty(0,\sin(\theta_3),\cos(\theta_3))^\intercal \vert i = 1, \dots, N)$
\end{itemize}

Figure \ref{fig:x3} shows the corresponding ensemble. The most probable states are close to the $y$-poles, which makes this ensemble distinct from the others.

\begin{table}[ht]
    \centering
    \begin{tabular}{c|cccc}
        $N$ & 1 & 2 & 3 & 4 \\
        \hline
        $t$ & 2.000 & 1.000 & 0.667 & 0.500 \\
        $g$ & 1.435 & 1.194 & 1.032 & 0.919 \\
        \hline
        $\theta_N$ & 1.570 & 0.748 & 0.456 & 0.334\\
    \end{tabular}
    \caption{Numerical parameters for the realization of our dephasing model with total time $T= Nt = 2.0$ split into different discretizations. $\theta_N$ is the polar angle characterizing Alice's third measurement strategy.}
    \label{tab:parameters}
\end{table}

\subsection{Measurements on Bob's side}
\label{sub:measurements-bob}
We have theoretically estimated which measurement scenarios Alice should implement to demonstrate steering. In each run she announces to Bob the measurement strategy $x$ she has implemented and the corresponding outcome $a$ she has obtained. In order to reconstruct the ensembles that Alice produces, Bob needs to implement an informationally complete (IC) set of measurements on his side to do quantum state tomography (QST). In general a single IC-POVM would suffice to acquire the necessary statistics. However, since such a POVM is always non-projective, an implementation on the IBMQ computer -- which can only measure projectively in the $\sigma_z$-basis --  would require additional ancillas. Thus, for experimental reasons it is more appropriate to measure a set of three projective POVMs $\Pi_i$
\begin{align}
\label{eq:ideal}
    & \mathcal{B}_\text{id} = \qty{\Pi_i}_{i=1}^3,\quad \textrm{with}\\
    & \Pi_i = \qty{B_i, \id-B_i} =  \qty{\frac{1}{2} \qty(\id \pm \sigma_i)}, 
\end{align}
which measure the different $\sigma_i$-components. Accordingly, for each state in the ensemble, Bob obtains a probability vector $\vec p_{a|x} = (p_1(0),p_2(0),p_3(0))^\intercal_{a|x}$, where $p_i(0)$ is the probability of the outcome ``0'' if POVM $\Pi_i$ was measured. Here we use the convention that outcome ``0'' refers to the POVM element $B_i = \frac{1}{2} \qty(\id + \sigma_i)$.

The POVMs in the set $\mathcal{B}_\text{id}$ describe perfect projective measurements. This is an idealization which is certainly not true in a real experiment. In other words, when Bob implements a $\Pi_i$-measurement, the POVM actually performed by the apparatus will slightly deviate from the ideal projective one. Quantum steering and QST, however, generally require a perfect characterization of the involved measurements. We will address this subtlety in detail in Sec.~\ref{sub:tomo} when we discuss the measured data.

\section{Implementation on IBMQ}
\label{sec:implementation}
\begin{figure}[ht]
    \centering
    \includegraphics[scale=0.9]{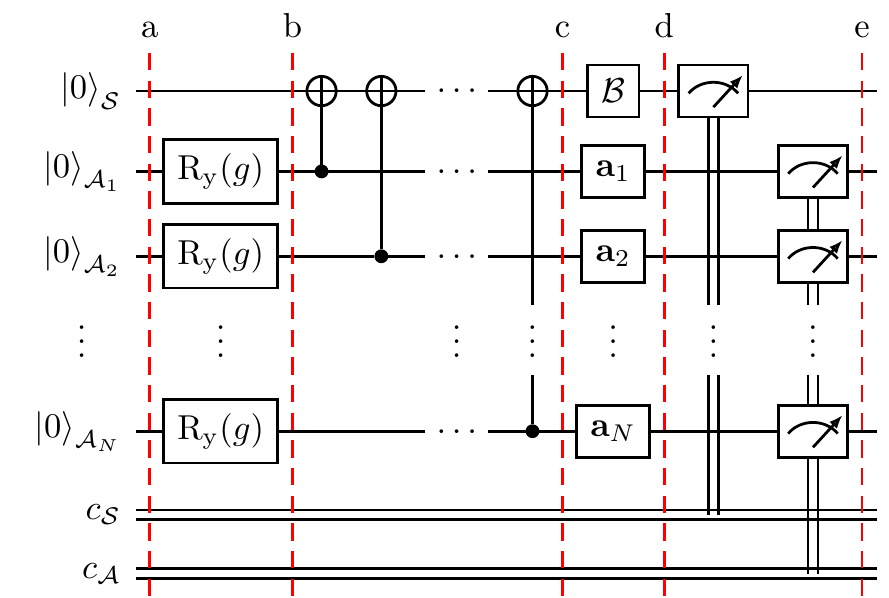}
    \caption{Fundamental quantum circuit for our dephasing process with $N$ collisions. Steps: (a) Preparing all qubits in the ground state. (b) Rotating the ancillas by parameter $g$. (c) Sequential application of ancilla-controlled CNOTs. (d) Measurement preparation by Alice and Bob. (e) Measurements in $\sigmaz$-basis, the outcomes for the system and the ancillas are stored in separate registers. }
    \label{fig:fund_circ}
\end{figure}

The fundamental quantum circuit for our dephasing model is shown in Figure \ref{fig:fund_circ}. All states are initially in the ground state $\ket{0}$ (a). First the ancillas are rotated according to the coupling parameter $g$ (b), which is followed by the sequential application of the \Op{CNOT} gates (c). Final measurements on IBMQ devices can only be conducted in the $\sigmaz$-basis (e). Hence Bob and Alice need to appropriately rotate their qubits beforehand in order to realize the desired measurements (d). Alice's measurement directions for the single ancilla are given by the strategies $x = (\vec{a}_1, \vec{a}_2, \dots, \vec{a}_N)$ derived in Sec.~\ref{sub:strats}. In each run one of the three measurements of the informationally complete set of POVMs $\mathcal{B}$ is implemented on Bob's side.

We implement our different collision models for $N=1,\dots,4$ on three quantum computers accessible through the IBM Quantum Experience cloud service \cite{ibmq2021}: 5-qubit devices \texttt{ibmq\_santiago} and \texttt{ibmq\_belem}, and 15-qubit device \texttt{ibmq\_16\_melbourne}. This makes twelve \emph{experiments}. Since these devices do not provide a star-like qubit topology, which would be preferred for collision models, inserting  \Op{SWAP} gates between collisions may be necessary. 
We conduct suitably many runs in order to collect a sufficient amount of data for performing state tomography for every state in the three-ensemble assemblage later. Further details on the implemented quantum circuits and on data acquisition can be found in Appendix \ref{app:implementation} and in Appendix \ref{app:data}, respectively. 

Before we continue with the analysis of the experimental data, we would like to point out that the given steering task cannot be loophole-free on an IBMQ device. First of all there is no random choice of measurement settings. Since the IBMQ machines can only measure in the computational basis, the different measurement directions are implemented by local unitary rotations before the read-out and, therefore, determined before each run. Accordingly, there is also clearly no space-like separation of both sides. However, a loophole-free steering task is not the aim of this paper and, as far as we see, not possible on current quantum computers. Thus, we rather benchmark to what extent the IBMQ machines are able to generate non-classical correlations necessary for semidevice-independent scenarios.

\section{Steering Analysis}
\label{sec:analysis}

\subsection{Quantum state tomography on Bob's side}
\label{sub:tomo}

The raw data for each scenario $x$ obtained from a single experiment consists of count statistics for the joined outcome space of Alice and Bob. We use Alice's $N$-bit outcome sequence $a$ to label each ensemble state, group Bob's POVM measurement outcomes in bins accordingly and estimate the probability vector $\vec{p}_{a|x}$. To reconstruct the ensemble states Bob performs QST for each of these vectors. Once the assemblage is reconstructed the steering weight can be computed with a semidefinite program as mentioned in Section \ref{sec:steering}.

The measurements on Bob's side are theoretically described by the idealized set of projective POVMs $\mathcal{B}_\text{id}$ in Eq.~(\ref{eq:ideal}). The actual measurements in the real experiment will not be perfectly projective. Therefore, we assume for the moment that the three measurements on Bob's side are given by general POVMs $\Pi_i = \qty{B_i,\id-B_i}$.
Each $B_i$ can be represented by a three-component vector $\vec b_i$ and a bias term $b^0_i$:
\begin{equation}
    \label{eq:POVM_element}
    B_i = \frac{1}{2} \qty(b_i^0 \id + \vec b_i \cdot \vec \sigma).
\end{equation}
We stress that the vector length $b_i = \abs{\vec{b}_i}$ and the bias term $b^0_i$ are constrained by the relation:
\begin{equation}
    \label{eq:constraint}
    b_i \leq b^0_i \leq 2-b_i \quad \Longleftrightarrow \quad b_i \leq \min \qty{b^0_i, 2-b^0_i}.
\end{equation}
This guarantees that the set $\Pi_i$ consists of positive operators and actually describes a POVM.

According to Born's rule the probability of measuring ``0'' when performing $\Pi_i$ on a qubit state with Bloch vector $\vec r$ is given by:
\begin{equation}
    \label{eq:meas_prob}
    p_i(0) = \frac{1}{2} \qty(b^0_i + \vec b_i \cdot \vec r) \qc i = 1,2,3.
\end{equation}
To reconstruct the Bloch vector $\vec r$ from the measured data $\vec p = (p_1(0),p_2(0),p_3(0))$ we have to solve this system of three linear equations:
\begin{equation}
    \vec r = {\underbrace{\mqty(\vec{b}_1^\intercal \\ \vec{b}_2^\intercal \\ \vec{b}_3^\intercal)}_{\equiv \mathbf{B}}}^{-1} \qty(2 \vec p - \vec{b^0}) = \mathbf{B}^{-1} \qty(2 \vec p - \vec{b^0}).
\end{equation}
Here $\vec{b^0} = \qty(b^0_1, b^0_2, b^0_3)$. The chosen set of POVMs is \emph{informationally complete} if $\mathbf{B}$ is invertible.
This QST approach is usually referred to as \emph{linear inversion} method \cite{GutaFast-state-tomography2020, DArianoOptimal-Data-Processing2007}.

Figure \ref{fig:ensembles} shows exemplary ensembles reconstructed from the $N=4$ experiment on \texttt{ibmq\_santiago}, depicted with blue dots. For this visualization we assumed in the tomography that the measurements were indeed given by the ideal POVM set $\mathcal{B} = \mathcal{B}_\text{id}$, so that it is comparable to the ideal theory. The blue triangles represent the ensemble averages, which should all be equal, and ideally, equivalent to the reduced state predicted by quantum theory (red triangles). This is only approximately true, which indicates that the no-signaling condition \eqref{eq:no-signaling} is slightly violated. This behavior is expected on a real machine due to typical crosstalk errors in quantum computers~\cite{SarovarDetectingcrosstalkerrors2020}.  

To verify quantum steering, we cannot simply assume the idealized measurements $\mathcal{B}_\textrm{id}$ in the matrix $\mathbf{B}$ of the QST. Instead we would need to \emph{perfectly} characterize the noise impacting the $\sigmaz$-measurement and the preceding qubit rotation on each device's physical qubits to obtain the actual set of measurements $\mathcal{B}$. This is of course impossible in a real experiment. Furthermore, the finite data measured in a QST can only approximate the underlying true probability distributions. As a consequence, the states obtained by the QST are in general not the actual states in the produced ensembles. This can most easily be seen by the well known fact that QST with the linear inversion method almost certainly predicts unphysical, i.e., non-positive density matrices. 
To cure these states, several techniques that enforce positive states have been proposed in the literature, such as maximum likelyhood methods~\cite{hradilQuantumstateEstimation1997,blume-kohoutHedgedMaximumLikelihood2010,smolinEfficientMethodComputing2012}, Bayesian approaches~\cite{Blume-KohoutOptimal-reliable-estimation2010,huszarAdaptiveBayesianQuantum2012,lukensPracticalEfficientApproach2020}, least square methods~\cite{opatrnyLeastsquaresInversionDensitymatrix1997,jamesMeasurementQubits2001} and linear regression estimation~\cite{qiQuantumStateTomography2013,houFullReconstruction14qubit2016,qiAdaptiveQuantumState2017}. For example, in the maximum likelihood method the non-positive states, which lie outside the Bloch sphere, are projected onto the surface, thereby altering the reduced state of the ensemble. This directly leads to a violation of the no-signaling condition \eqref{eq:no-signaling}, which is problematic in a steering scenario. Furthermore, the purity of the states predicted by the tomography is then often overestimated. While this is usually not a big issue for local observables, it severely influences the estimation of device-dependent non-local properties such as entanglement. For instance, in Ref.~\cite{schwemmerSystematicErrorsCurrent2015} it has been shown that maximum likelihood and least square methods systematically overestimate the entanglement in the measured state. Recently a method that allows to verify steering without full characterization of the measurement apparatus has been proposed~\cite{zhaoExperimentalDemonstrationMeasurementdeviceindependent2020}. However, this approach relies on a source of trusted quantum states and is therefore also not applicable to our experimental setup. 

To circumvent these problems we make use of a more practical approach. Strictly speaking we are not interested in the quantum states on Bob's side but these are only a means of demonstrating steering. Therefore, we stick to linear inversion but do not rely on a specific reconstruction of Bob's states. Instead we search for a lower bound \Op{LB} for every assemblage, which we realize by minimizing the steering weight over all \emph{valid} sets of POVMs $\mathcal{B}$ that determine the state tomography:
\begin{equation}
    \Op{LB} = \min_\mathcal{B} \Op{SW} \quad \text{s.t.} \quad \mathcal{B} \text{ is \emph{valid}}.
\end{equation}
Here we say that a set of POVMs $B$ is \emph{valid} if all assemblage states obtained from the tomography are valid quantum states. In particular, assuming for example compatible POVMs on Bob's side (a choice which would not be able to show steering), the tomography would in general produce non-positive states. Thus, the assumption of such POVMs is incompatible with the measured data.

Bob's POVMs are described by three three-component vectors $\vec{b}_i$ and three bias terms $b^0_i$, making a total of twelve parameters. However, we do not need to distinguish between all of these sets. The definition of the steering weight \eqref{eq:SW} suggests a unitary freedom of the input assemblage, which translates to a unitary freedom of the tomography POVM set due to Born's rule. This again translates to a rotational freedom of the vectors ${\vec{b}_i} \in\mathbb{R}^3$. Therefore, w.l.o.g., we can set
\begin{equation}
\begin{split}
    \vec{b}_3 &= b_3 \mqty(0 \\ 0 \\ 1) \qc \vec{b}_1 = b_1 \mqty(\sin(\theta_1) \\ 0 \\ \cos(\theta_1)), \\
    \vec{b}_2 &= b_2 \mqty(-\sin(\varphi_2) \sin(\theta_2) \\ \cos(\varphi_2) \sin(\theta_2) \\ \cos(\theta_2)).
\end{split}
\end{equation}
Here $\theta_1, \theta_2 \in (0, \pi)$ are the usual polar angles, while the azimuthal angle $\varphi_2 \in (-\pi, \pi)$ is measured from the $y$-axis. The ideal POVM set $\mathcal{B}_\text{id}$ translates to $\theta_{1, \text{id}} = \theta_{2, \text{id}} = \nicefrac{\pi}{2}$ and $\varphi_{2, \text{id}} = 0$. Thus we are left with nine parameters and our minimization problem becomes \footnote{The validity constraint is implied and will not be noted henceforth.}:
\begin{equation}
    \Op{LB}= \min_{b^0_i, b_i, \theta_1, \theta_2, \varphi_2} \Op{SW} \quad \text{s.t.} \quad \text{constraint \eqref{eq:constraint} holds}.
\end{equation}
This minimization problem is not convex. For the small assemblages ($N=1,2$) we compute the lower bound $\Op{LB}$ using several global minimization algorithms as well as a gradient descent approach for multiple initial states. As all these methods converge to the same minimum we consider the problem well-behaved. 
Furthermore, we find that the minima always correspond to projective measurements, i.e. $b_i=b^0_i=1$. 
This is reasonable, because vector lengths $b_i < 1$ in effect increase state purity through the tomography, which causes an increase of the steering weight.
The measurements on the IBMQ devices are certainly not perfectly projective. Moreover, due to quantum decay, one generally observes a bias towards the ``0'' outcome. However, the full characteristics of the performed measurements are unknown and, therefore, assuming projectors ensures not to overestimate the steerability.
Then the final form of our minimization problem comprises only three parameters 
(the relative angles between the measurement directions)
and simply reads:
\begin{equation}
    \Op{LB}= \min_{\theta_1, \theta_2, \varphi_2} \Op{SW}.
\end{equation}
For the larger data sets of $N=3,4$ collisions we then stick to the gradient descent method in order to reduce the number of calls of the expensive steering weight SDP, which allows us to process the data on a conventional workstation in reasonable time.

\subsection{Results}
\begin{figure}[t]
    \centering
    \includegraphics[width=\linewidth]{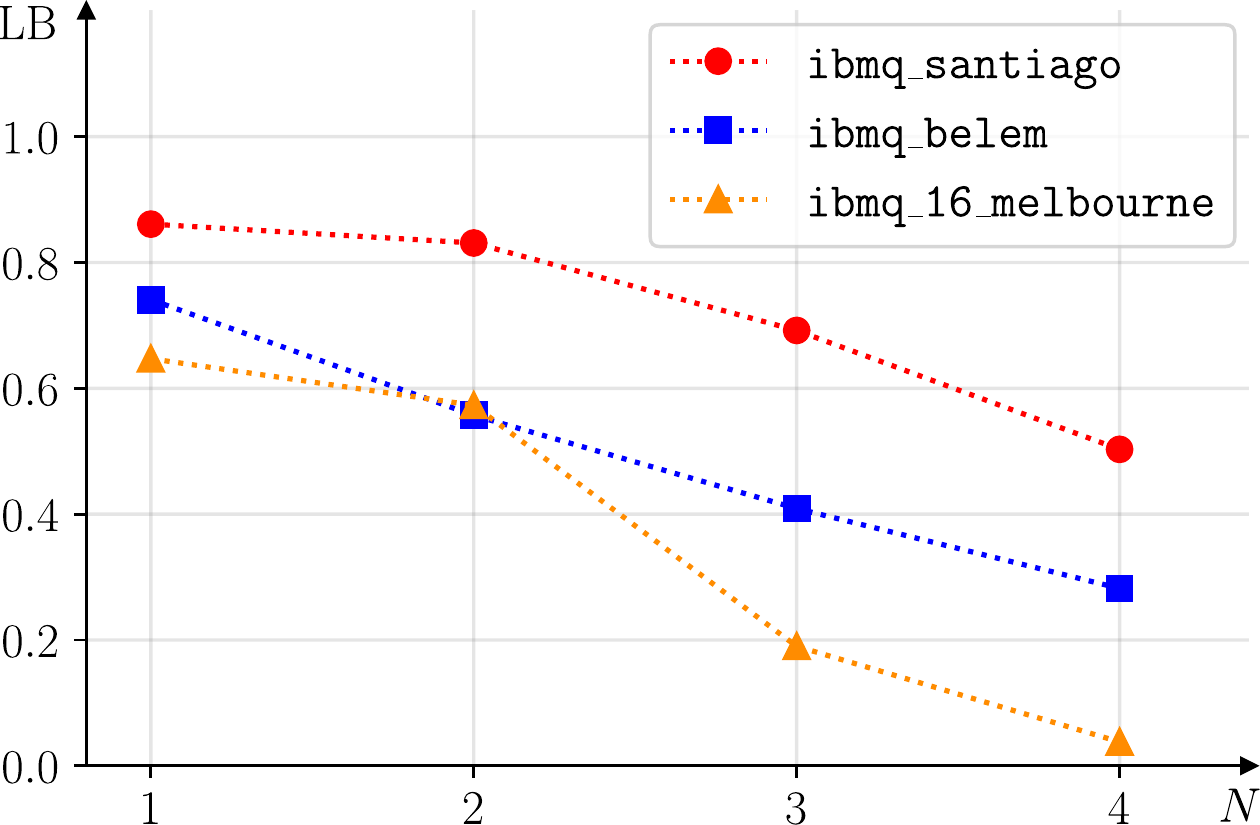}
    \caption{Steering weight lower bounds $\Op{LB}$ from experiments with different numbers of collisions $N$. Increasing experiment complexity causes the decline of quantum correlations on all devices. Data also available in Table \ref{tab:results}.}
    \label{fig:LB}
\end{figure}

We visualize the minimization results in Figure \ref{fig:LB} and present the numbers in Table \ref{tab:results}. Interestingly, \texttt{ibmq\_santiago} produces the largest steering weight in all cases, despite providing the most unfavorable qubit layout. This indicates that this device is affected by the least noise, which coincides with IBM reporting it to have the largest quantum volume out of our selection \cite{ibmq2021}. Still, the impact of the necessary \Op{SWAP} operation is apparent when going from two to three collisions. here the lower bound of the steering weight drops significantly, as does when going from three to four collisions. 

This effect is even more prominent when adding the first \Op{SWAP} for the 3-collisions experiment on \texttt{ibmq\_16\_melbourne}. This device performs poorly in the case of four collisions as there is barely any detectable steering left.

\texttt{ibmq\_belem} presents itself as the average device of our selection, with a decent amount of steering left even after four collisions. Unfortunately, as with \texttt{ibmq\_santiago}, we cannot increase the number of collisions on this machine as there are not enough physical qubits available. While \texttt{ibmq\_16\_melbourne} would in principle allow up to 15 collisions, the noise impact in this machine prevents us from detecting steering for more than four collisions.  

We point out that the minimizing POVM sets deviate from the ideal case by only a few degrees in each angle. However, keeping in mind the freedom of a global unitary rotation on the set of POVMs, these angles merely describe the relative position of the POVM elements with respect to each other. 
\begin{table*}[ht]
    \centering
    \begin{tabular}{c||ccc|c||ccc|c||ccc|c||ccc|c}
        $N$ & \multicolumn{4}{c||}{1} & \multicolumn{4}{c||}{2} & \multicolumn{4}{c||}{3} & \multicolumn{4}{c}{4} \\
        \hline
        $\min$ & $\theta_1 [^\circ]$ & $\theta_2 [^\circ]$ & $\varphi_2 [^\circ]$ & \Op{LB} & $\theta_1 [^\circ]$ & $\theta_2 [^\circ]$ & $\varphi_2 [^\circ]$ & \Op{LB} & $\theta_1 [^\circ]$ & $\theta_2 [^\circ]$ & $\varphi_2 [^\circ]$ & \Op{LB} & $\theta_1 [^\circ]$ & $\theta_2 [^\circ]$ & $\varphi_2 [^\circ]$ & \Op{LB} \\
        \hline
        \texttt{ibmq\_santiago} & 87.69 & 90.35 & $-0.17$ & 0.861 & 86.62 & 93.16 & $-1.33$ & 0.831 & 85.30 & 92.18 & $-1.11$ & 0.692 & 86.16 & 91.78 & 0.04 & 0.503 \\
        \texttt{ibmq\_belem} & 87.90 & 87.61 & 0.18 & 0.741 & 85.14 & 90.26 & $-1.26$ & 0.557 & 82.67 & 93.36 & $-5.05$ & 0.409 & 82.55 & 93.45 & $-5.65$ & 0.282 \\
        \texttt{ibmq\_16\_melbourne} & 86.33 & 87.13 & $-0.65$ & 0.647 & 84.20 & 90.95 & $-2.31$ & 0.573 & 84.29 & 90.58 & $-9.28$ & 0.190 & 84.64 & 89.89 & 3.71 & 0.038
    \end{tabular}
    \caption{Steering weight lower bounds $\Op{LB}$ and angles parametrizing the projective POVM sets, which realize these lower bounds. Each of them is a few degrees away from the ideal case $\mathcal{B}_{\text{id}}$ (up to global unitary rotations). We find presence of steering in all experiments, although with different strengths. In accordance with IBM's reports, we conclude that \texttt{ibmq\_santiago} is the best performing device from our selection, while \texttt{ibmq\_16\_melbourne} performs worst.} 
    \label{tab:results}
\end{table*}

The analysis of the steerability of Bob's system by Alice's measurements is of course motivated by the asymmetric scenario of system and environment. Nevertheless, one might wonder if the measured data could also violate a Bell inequality. For our measured data the answer is negative which is due to the fact that the angles between the measurement directions on Bob's side are approximately $90^\circ$. For the violation of the CHSH inquality, for example, one would choose an angle of $45^\circ$~\cite{clauserProposedExperimentTest1969}. Nevertheless, using the measured data we can estimate what Bob would have obtained if he had measured with different angles. These estimations -- which are of course device-dependent and therefore do not replace a Bell test -- suggest that the scenario could also violate a Bell inequality, at least on the best device \texttt{ibmq\_santiago}. 

\section{Conclusion}
We successfully showed the presence of quantum steering between an open quantum system and its environment in different settings on contemporary quantum computers. After having reviewed the fundamentals of quantum steering, we presented a specific dephasing model as a promising candidate for investigating quantum steering in a collision model setup on real quantum machines. We considered a process of constant time, i.e., a fixed dephasing channel, induced by an increasing number of collisions $N=1, \dots, 4$, being interested in the impact on the chosen steering quantifier. Since we initialized system and environment in a product state, we can be certain that the observed quantum correlations emerged from the collisions, proving that the dephasing channel was induced by the quantum nature of the environment.

We employed three different measurement strategies on Alice's environment side to produce an assemblage on Bob's system side. Running simulations on the three IBMQ quantum computers \texttt{ibmq\_santiago}, \texttt{ibmq\_belem} and \texttt{ibmq\_16\_melbourne} revealed their different perfomance qualities and noise levels. We chose the steering weight as our steering quantifier. It is computed with the help of a semidefinite program that takes an experimental assemblage as its input, once it is determined from state tomography on Bob's end. In order to eliminate the necessity of knowing Bob's measurement apparatus perfectly, we minimized the steering weight over all valid sets of POVMs for every experiment. We found that the minimizing measurements are projective but do not correspond to the ideal case of measuring in three mutually orthogonal directions on the Bloch sphere.  
The resulting lower bounds provide insight into the quality of each quantum device and show the impact of more collisions on the steering quantifier.

As can be seen in Table \ref{tab:results}, 5-qubit device \texttt{ibmq\_santiago} performed best from our selection, showing the greatest steering weights for all collision numbers. This coincides with IBM's reported quantum volumes, a certain metric for a quantum computer's quality \cite{ibmq2021}. The other 5-qubit machine \texttt{ibmq\_belem} produced intermediate results, showing a significant loss in the steering weight after $N=4$ collisions. Even worse performance can be observed for the larger 15-qubit device \texttt{ibmq\_16\_melbourne}, with barely any detectable steering left in this case.

We observed small violations of the no-signaling condition \eqref{eq:no-signaling} as the ensemble averages of a given experiment slightly differ from each other (see Figure \ref{fig:ensembles}). Strictly speaking this contests the entire steering concept. However, we are certain that this effect cannot be circumvented, because it is caused by the unavoidable imperfections real quantum computers will always bear. Unpredictable erroneous gates and measurements will not allow perfect, reproducible simulations of quantum processes. Therefore, we still talk of quantum steering, as our approach represents the closest we can get to the ideal case.

Expanding our approach to more collisions, which would justify talking of ``open quantum system dynamics'' even more, would be of great interest. \texttt{ibmq\_16\_melbourne} would allow such an endeavor, but our results show that the chances of observing steering would be minimal. 

\begin{acknowledgments}
We acknowledge the use of IBM Quantum services for this work. The views expressed are those of the authors, and do not reflect the official policy or position of IBM or the IBM Quantum team. We would like to thank Roope Uola for useful discussions.
\end{acknowledgments}

\appendix

\section{Implementing the quantum circuits}
\label{app:implementation}
\subsection{Realizing Bob's measurements}
The Pauli matrices are a natural choice for suitable observables for state tomography. Thus the ideal set of POVMs we would wish to implement is $\mathcal{B}_\text{id} = \qty{\Pi_i = \qty{\nicefrac{1}{2} \qty(\id \pm \sigma_i)}}_{i=1}^3$. Therefore, if Bob wants to measure $\sigmax$ or $\sigmay$, he needs to apply the appropriate gate just before the $\sigmaz$-measurement as shown in Figure \ref{fig:fund_circ}:
\begin{equation}
    \sigmax \xrightarrow{\text{apply}} \Op{R_y}\qty(-\frac{\pi}{2}) \qc \sigmay \xrightarrow{\text{apply}} \Op{R_x}\qty(\frac{\pi}{2}).
\end{equation}

In reality though, these operations will not be free of errors, which is why the true POVM set $\mathcal{B}$ remains unknown. With our analysis method described above we eliminate the necessity of having this knowledge, but we still assume that the POVM elements do not change over time and are constant for the entirety of an experiment.

\subsection{Realizing Alice's measurements}
In order to implement the different measurement scenarios presented in Section \ref{sub:strats}, Alice may perform additional manipulations on all her ancillas prior to the final measurements as shown in Figure \ref{fig:fund_circ}. Strategy $x_1$ simply consists of local $\sigmaz$-measurements, so this extra step is not necessary. However, strategies $x_2$ and $x_3$ require the following operations:
\begin{equation}
    x_2 \xrightarrow{\text{apply}} \Op{R_y}(-g) \qc x_3 \xrightarrow{\text{apply}} \Op{R_x}\qty(\theta_N).
\end{equation}

\subsection{Circuit mapping and \Op{SWAP} operations}
\begin{figure*}[ht]
    \centering
    \includegraphics[scale=0.9]{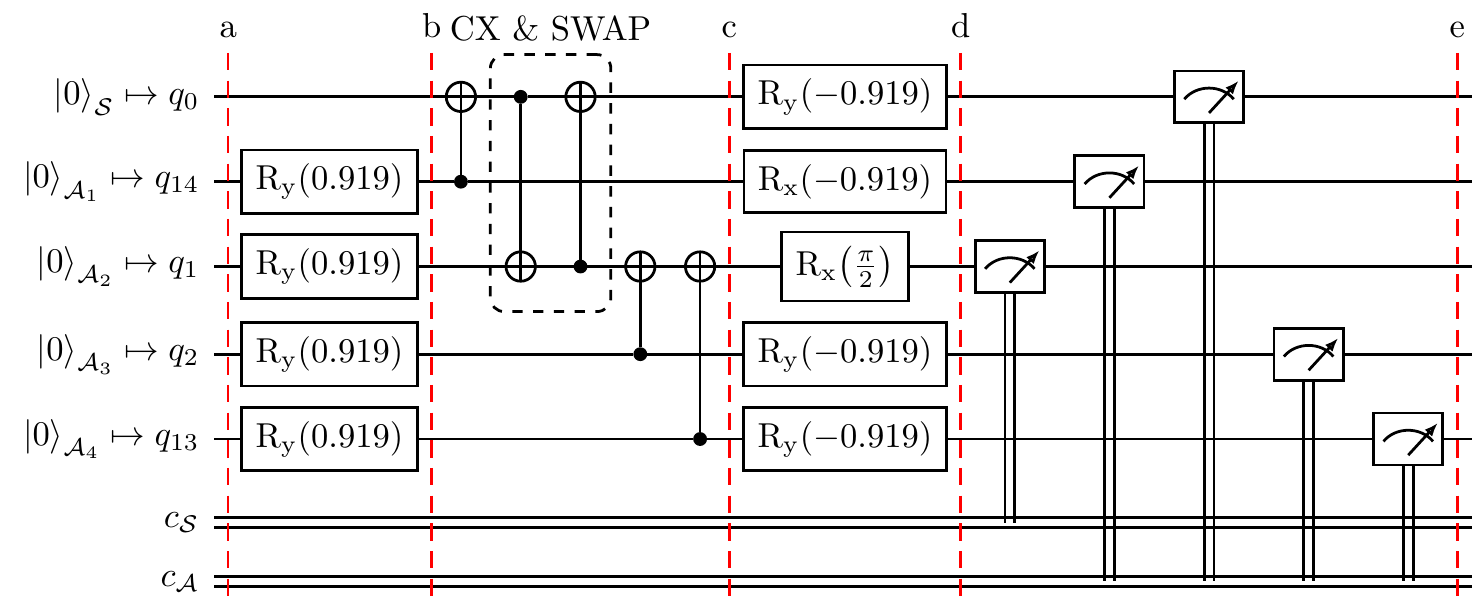}
    \caption{Quantum circuit like it was implemented on \texttt{ibmq\_16\_melbourne} as part of a 4-collision experiment on 11/25/2020. Steps: (a) Mapping the logical qubits  onto the device's physical qubits. (b) Rotating the ancillas according to the coupling parameter $g=0.919$. (c) Applying the \Op{CNOT} gates. One \Op{CNOT} is contracted with the necessary \Op{SWAP} operation as highlighted by the dashed box. (d) Alice prepares to realize scenario $x_2$, Bob prepares to measure $\sigmay$. (e) Final $\sigmaz$-measurements.}
    \label{fig:real_circ}
\end{figure*}

A crucial part of quantum computing influencing its performance is the mapping of a given quantum circuit with logical qubits onto the physical qubits of a quantum device. We achieve this by studying the noise maps provided by IBM and assigning the qubits ``by hand''. This procedure could be further optimized. We directly include the suitable positioning of \Op{SWAP} gates into our consideration, which is partially necessary because none of the selected device offers a star-like qubit layout. Note that our specific model allows the neat contraction of a \Op{CNOT} gate and a costly \Op{SWAP} gate into two \Op{CNOT} gates. An exemplary quantum circuit is shown in Figure~\ref{fig:real_circ}.

\section{Data acquisition}
\label{app:data}
A single experiment consists of 9 different quantum circuits, one for each combination of Alice's strategies $x$ and Bob's observables $\Pi_i$. Every circuit needs to be executed multiple times so that we obtain well-converged measurement statistics. With an increasing number of collisions $N$ the amount of states forming an assemblage grows exponentially and thus we require an exponentially increasing number of circuit runs. IBM allows the execution of 8192 shots per quantum circuit and 75 circuits per submitted job. We design a base job, which contains 8 copies of each experiment circuit, and submit it repeatedly. Table \ref{tab:shots} shows the numbers. Data recording took place on 11/23/2020-11/25/2020 (\texttt{ibmq\_santiago}, \texttt{ibmq\_16\_melbourne}) and 03/10/2021-03/11/2021 (\texttt{ibmq\_belem}).
\begin{table}[ht]
    \centering
    \begin{tabular}{c|cccc}
        $N$ & 1 & 2 & 3 & 4 \\
        \hline
        job repetitions & 10 & 16 & 30 & 60 \\
        shots/circuit & 655,360 & 1,048,576 & 1,966,080 & 3,932,160
    \end{tabular}
    \caption{Number of executions for every unique experiment circuit depending on the number of collisions $N$, where $\text{shots/circuit} = (\text{job repetitions}) \cdot 8 \cdot 8192.$}
    \label{tab:shots}
\end{table}

\bibliography{ibmq.bib}

\begin{thebibliography}{54}%
\makeatletter
\providecommand \@ifxundefined [1]{%
 \@ifx{#1\undefined}
}%
\providecommand \@ifnum [1]{%
 \ifnum #1\expandafter \@firstoftwo
 \else \expandafter \@secondoftwo
 \fi
}%
\providecommand \@ifx [1]{%
 \ifx #1\expandafter \@firstoftwo
 \else \expandafter \@secondoftwo
 \fi
}%
\providecommand \natexlab [1]{#1}%
\providecommand \enquote  [1]{``#1''}%
\providecommand \bibnamefont  [1]{#1}%
\providecommand \bibfnamefont [1]{#1}%
\providecommand \citenamefont [1]{#1}%
\providecommand \href@noop [0]{\@secondoftwo}%
\providecommand \href [0]{\begingroup \@sanitize@url \@href}%
\providecommand \@href[1]{\@@startlink{#1}\@@href}%
\providecommand \@@href[1]{\endgroup#1\@@endlink}%
\providecommand \@sanitize@url [0]{\catcode `\\12\catcode `\$12\catcode
  `\&12\catcode `\#12\catcode `\^12\catcode `\_12\catcode `\%12\relax}%
\providecommand \@@startlink[1]{}%
\providecommand \@@endlink[0]{}%
\providecommand \url  [0]{\begingroup\@sanitize@url \@url }%
\providecommand \@url [1]{\endgroup\@href {#1}{\urlprefix }}%
\providecommand \urlprefix  [0]{URL }%
\providecommand \Eprint [0]{\href }%
\providecommand \doibase [0]{http://dx.doi.org/}%
\providecommand \selectlanguage [0]{\@gobble}%
\providecommand \bibinfo  [0]{\@secondoftwo}%
\providecommand \bibfield  [0]{\@secondoftwo}%
\providecommand \translation [1]{[#1]}%
\providecommand \BibitemOpen [0]{}%
\providecommand \bibitemStop [0]{}%
\providecommand \bibitemNoStop [0]{.\EOS\space}%
\providecommand \EOS [0]{\spacefactor3000\relax}%
\providecommand \BibitemShut  [1]{\csname bibitem#1\endcsname}%
\let\auto@bib@innerbib\@empty
\bibitem [{\citenamefont {Einstein}\ \emph {et~al.}(1935)\citenamefont
  {Einstein}, \citenamefont {Podolsky},\ and\ \citenamefont
  {Rosen}}]{EinsteinCan-Quantum-Mechanical-Description-of-Physical1935}%
  \BibitemOpen
  \bibfield  {author} {\bibinfo {author} {\bibfnamefont {A.}~\bibnamefont
  {Einstein}}, \bibinfo {author} {\bibfnamefont {B.}~\bibnamefont {Podolsky}},
  \ and\ \bibinfo {author} {\bibfnamefont {N.}~\bibnamefont {Rosen}},\ }\href
  {\doibase 10.1103/PhysRev.47.777} {\bibfield  {journal} {\bibinfo  {journal}
  {Phys. Rev.}\ }\textbf {\bibinfo {volume} {47}},\ \bibinfo {pages} {777}
  (\bibinfo {year} {1935})}\BibitemShut {NoStop}%
\bibitem [{\citenamefont
  {Schr{\"o}dinger}(1935)}]{SchrodingerDiscussion-of-Probability-Relations1935}%
  \BibitemOpen
  \bibfield  {author} {\bibinfo {author} {\bibfnamefont {E.}~\bibnamefont
  {Schr{\"o}dinger}},\ }\bibfield  {booktitle} {\emph {\bibinfo {booktitle}
  {Mathematical Proceedings of the Cambridge Philosophical Society}},\ }\href
  {\doibase DOI: 10.1017/S0305004100013554} {\ \textbf {\bibinfo {volume}
  {31}},\ \bibinfo {pages} {555} (\bibinfo {year} {1935})}\BibitemShut
  {NoStop}%
\bibitem [{\citenamefont {Wiseman}\ \emph {et~al.}(2007)\citenamefont
  {Wiseman}, \citenamefont {Jones},\ and\ \citenamefont
  {Doherty}}]{wisemanSteeringEntanglementNonlocality2007}%
  \BibitemOpen
  \bibfield  {author} {\bibinfo {author} {\bibfnamefont {H.~M.}\ \bibnamefont
  {Wiseman}}, \bibinfo {author} {\bibfnamefont {S.~J.}\ \bibnamefont {Jones}},
  \ and\ \bibinfo {author} {\bibfnamefont {A.~C.}\ \bibnamefont {Doherty}},\
  }\href {\doibase 10.1103/PhysRevLett.98.140402} {\bibfield  {journal}
  {\bibinfo  {journal} {Physical Review Letters}\ }\textbf {\bibinfo {volume}
  {98}} (\bibinfo {year} {2007}),\ 10.1103/PhysRevLett.98.140402}\BibitemShut
  {NoStop}%
\bibitem [{\citenamefont {Jones}\ \emph {et~al.}(2007)\citenamefont {Jones},
  \citenamefont {Wiseman},\ and\ \citenamefont
  {Doherty}}]{jonesEntanglementEinsteinPodolskyRosenCorrelations2007}%
  \BibitemOpen
  \bibfield  {author} {\bibinfo {author} {\bibfnamefont {S.~J.}\ \bibnamefont
  {Jones}}, \bibinfo {author} {\bibfnamefont {H.~M.}\ \bibnamefont {Wiseman}},
  \ and\ \bibinfo {author} {\bibfnamefont {A.~C.}\ \bibnamefont {Doherty}},\
  }\href {\doibase 10.1103/PhysRevA.76.052116} {\bibfield  {journal} {\bibinfo
  {journal} {Physical Review A}\ }\textbf {\bibinfo {volume} {76}} (\bibinfo
  {year} {2007}),\ 10.1103/PhysRevA.76.052116}\BibitemShut {NoStop}%
\bibitem [{\citenamefont {Uola}\ \emph {et~al.}(2020)\citenamefont {Uola},
  \citenamefont {Costa}, \citenamefont {Nguyen},\ and\ \citenamefont
  {G{\"u}hne}}]{uolaQuantumSteering2020}%
  \BibitemOpen
  \bibfield  {author} {\bibinfo {author} {\bibfnamefont {R.}~\bibnamefont
  {Uola}}, \bibinfo {author} {\bibfnamefont {A.~C.~S.}\ \bibnamefont {Costa}},
  \bibinfo {author} {\bibfnamefont {H.~C.}\ \bibnamefont {Nguyen}}, \ and\
  \bibinfo {author} {\bibfnamefont {O.}~\bibnamefont {G{\"u}hne}},\ }\href
  {\doibase 10.1103/RevModPhys.92.015001} {\bibfield  {journal} {\bibinfo
  {journal} {Reviews of Modern Physics}\ }\textbf {\bibinfo {volume} {92}},\
  \bibinfo {pages} {015001} (\bibinfo {year} {2020})}\BibitemShut {NoStop}%
\bibitem [{\citenamefont {Cavalcanti}\ \emph {et~al.}(2009)\citenamefont
  {Cavalcanti}, \citenamefont {Jones}, \citenamefont {Wiseman},\ and\
  \citenamefont {Reid}}]{CavalcantiExperimental-criteria-for-steering2009}%
  \BibitemOpen
  \bibfield  {author} {\bibinfo {author} {\bibfnamefont {E.~G.}\ \bibnamefont
  {Cavalcanti}}, \bibinfo {author} {\bibfnamefont {S.~J.}\ \bibnamefont
  {Jones}}, \bibinfo {author} {\bibfnamefont {H.~M.}\ \bibnamefont {Wiseman}},
  \ and\ \bibinfo {author} {\bibfnamefont {M.~D.}\ \bibnamefont {Reid}},\
  }\href {\doibase 10.1103/PhysRevA.80.032112} {\bibfield  {journal} {\bibinfo
  {journal} {Phys. Rev. A}\ }\textbf {\bibinfo {volume} {80}},\ \bibinfo
  {pages} {032112} (\bibinfo {year} {2009})}\BibitemShut {NoStop}%
\bibitem [{\citenamefont {Skrzypczyk}\ \emph {et~al.}(2014)\citenamefont
  {Skrzypczyk}, \citenamefont {Navascu\'es},\ and\ \citenamefont
  {Cavalcanti}}]{SkrzypczykQuantifying-Einstein-Podolsky-Rosen-Steering2014}%
  \BibitemOpen
  \bibfield  {author} {\bibinfo {author} {\bibfnamefont {P.}~\bibnamefont
  {Skrzypczyk}}, \bibinfo {author} {\bibfnamefont {M.}~\bibnamefont
  {Navascu\'es}}, \ and\ \bibinfo {author} {\bibfnamefont {D.}~\bibnamefont
  {Cavalcanti}},\ }\href {\doibase 10.1103/PhysRevLett.112.180404} {\bibfield
  {journal} {\bibinfo  {journal} {Phys. Rev. Lett.}\ }\textbf {\bibinfo
  {volume} {112}},\ \bibinfo {pages} {180404} (\bibinfo {year}
  {2014})}\BibitemShut {NoStop}%
\bibitem [{\citenamefont {Quintino}\ \emph {et~al.}(2015)\citenamefont
  {Quintino}, \citenamefont {V\'ertesi}, \citenamefont {Cavalcanti},
  \citenamefont {Augusiak}, \citenamefont {Demianowicz}, \citenamefont
  {Ac\'{\i}n},\ and\ \citenamefont
  {Brunner}}]{QuintinoInequivalence-of-entanglement-steering2015}%
  \BibitemOpen
  \bibfield  {author} {\bibinfo {author} {\bibfnamefont {M.~T.}\ \bibnamefont
  {Quintino}}, \bibinfo {author} {\bibfnamefont {T.}~\bibnamefont {V\'ertesi}},
  \bibinfo {author} {\bibfnamefont {D.}~\bibnamefont {Cavalcanti}}, \bibinfo
  {author} {\bibfnamefont {R.}~\bibnamefont {Augusiak}}, \bibinfo {author}
  {\bibfnamefont {M.}~\bibnamefont {Demianowicz}}, \bibinfo {author}
  {\bibfnamefont {A.}~\bibnamefont {Ac\'{\i}n}}, \ and\ \bibinfo {author}
  {\bibfnamefont {N.}~\bibnamefont {Brunner}},\ }\href {\doibase
  10.1103/PhysRevA.92.032107} {\bibfield  {journal} {\bibinfo  {journal} {Phys.
  Rev. A}\ }\textbf {\bibinfo {volume} {92}},\ \bibinfo {pages} {032107}
  (\bibinfo {year} {2015})}\BibitemShut {NoStop}%
\bibitem [{\citenamefont {Cavalcanti}\ and\ \citenamefont
  {Skrzypczyk}(2017)}]{cavalcantiQuantumSteeringReview2017}%
  \BibitemOpen
  \bibfield  {author} {\bibinfo {author} {\bibfnamefont {D.}~\bibnamefont
  {Cavalcanti}}\ and\ \bibinfo {author} {\bibfnamefont {P.}~\bibnamefont
  {Skrzypczyk}},\ }\href {\doibase 10.1088/1361-6633/80/2/024001} {\bibfield
  {journal} {\bibinfo  {journal} {Reports on Progress in Physics}\ }\textbf
  {\bibinfo {volume} {80}},\ \bibinfo {pages} {024001} (\bibinfo {year}
  {2017})}\BibitemShut {NoStop}%
\bibitem [{\citenamefont {Wollmann}\ \emph {et~al.}(2018)\citenamefont
  {Wollmann}, \citenamefont {Hall}, \citenamefont {Patel}, \citenamefont
  {Wiseman},\ and\ \citenamefont
  {Pryde}}]{WollmannReference-frame-independent-Einstein-Podolsky-Rosen-steering2018}%
  \BibitemOpen
  \bibfield  {author} {\bibinfo {author} {\bibfnamefont {S.}~\bibnamefont
  {Wollmann}}, \bibinfo {author} {\bibfnamefont {M.~J.~W.}\ \bibnamefont
  {Hall}}, \bibinfo {author} {\bibfnamefont {R.~B.}\ \bibnamefont {Patel}},
  \bibinfo {author} {\bibfnamefont {H.~M.}\ \bibnamefont {Wiseman}}, \ and\
  \bibinfo {author} {\bibfnamefont {G.~J.}\ \bibnamefont {Pryde}},\ }\href
  {\doibase 10.1103/PhysRevA.98.022333} {\bibfield  {journal} {\bibinfo
  {journal} {Phys. Rev. A}\ }\textbf {\bibinfo {volume} {98}},\ \bibinfo
  {pages} {022333} (\bibinfo {year} {2018})}\BibitemShut {NoStop}%
\bibitem [{\citenamefont {Beyer}\ \emph {et~al.}(2019)\citenamefont {Beyer},
  \citenamefont {Luoma},\ and\ \citenamefont
  {Strunz}}]{beyerSteeringHeatEngines2019}%
  \BibitemOpen
  \bibfield  {author} {\bibinfo {author} {\bibfnamefont {K.}~\bibnamefont
  {Beyer}}, \bibinfo {author} {\bibfnamefont {K.}~\bibnamefont {Luoma}}, \ and\
  \bibinfo {author} {\bibfnamefont {W.~T.}\ \bibnamefont {Strunz}},\ }\href
  {\doibase 10.1103/PhysRevLett.123.250606} {\bibfield  {journal} {\bibinfo
  {journal} {Physical Review Letters}\ }\textbf {\bibinfo {volume} {123}},\
  \bibinfo {pages} {250606} (\bibinfo {year} {2019})}\BibitemShut {NoStop}%
\bibitem [{\citenamefont {Zheng}\ \emph {et~al.}(2020)\citenamefont {Zheng},
  \citenamefont {Sun}, \citenamefont {Yuan}, \citenamefont {Ficek},
  \citenamefont {Gong},\ and\ \citenamefont
  {He}}]{zhengEnhancedEntanglementAsymmetric2020}%
  \BibitemOpen
  \bibfield  {author} {\bibinfo {author} {\bibfnamefont {S.-S.}\ \bibnamefont
  {Zheng}}, \bibinfo {author} {\bibfnamefont {F.-X.}\ \bibnamefont {Sun}},
  \bibinfo {author} {\bibfnamefont {H.-Y.}\ \bibnamefont {Yuan}}, \bibinfo
  {author} {\bibfnamefont {Z.}~\bibnamefont {Ficek}}, \bibinfo {author}
  {\bibfnamefont {Q.-H.}\ \bibnamefont {Gong}}, \ and\ \bibinfo {author}
  {\bibfnamefont {Q.-Y.}\ \bibnamefont {He}},\ }\href {\doibase
  10.1007/s11433-020-1587-5} {\bibfield  {journal} {\bibinfo  {journal}
  {Science China Physics, Mechanics \& Astronomy}\ }\textbf {\bibinfo {volume}
  {64}},\ \bibinfo {pages} {210311} (\bibinfo {year} {2020})}\BibitemShut
  {NoStop}%
\bibitem [{\citenamefont {Designolle}\ \emph {et~al.}(2021)\citenamefont
  {Designolle}, \citenamefont {Srivastav}, \citenamefont {Uola}, \citenamefont
  {Valencia}, \citenamefont {McCutcheon}, \citenamefont {Malik},\ and\
  \citenamefont {Brunner}}]{designolleGenuineHighDimensionalQuantum2021}%
  \BibitemOpen
  \bibfield  {author} {\bibinfo {author} {\bibfnamefont {S.}~\bibnamefont
  {Designolle}}, \bibinfo {author} {\bibfnamefont {V.}~\bibnamefont
  {Srivastav}}, \bibinfo {author} {\bibfnamefont {R.}~\bibnamefont {Uola}},
  \bibinfo {author} {\bibfnamefont {N.~H.}\ \bibnamefont {Valencia}}, \bibinfo
  {author} {\bibfnamefont {W.}~\bibnamefont {McCutcheon}}, \bibinfo {author}
  {\bibfnamefont {M.}~\bibnamefont {Malik}}, \ and\ \bibinfo {author}
  {\bibfnamefont {N.}~\bibnamefont {Brunner}},\ }\href {\doibase
  10.1103/PhysRevLett.126.200404} {\bibfield  {journal} {\bibinfo  {journal}
  {Physical Review Letters}\ }\textbf {\bibinfo {volume} {126}},\ \bibinfo
  {pages} {200404} (\bibinfo {year} {2021})}\BibitemShut {NoStop}%
\bibitem [{\citenamefont {Pan}\ \emph {et~al.}(2021)\citenamefont {Pan},
  \citenamefont {Yang}, \citenamefont {Yuan}, \citenamefont {Zhang},\ and\
  \citenamefont {Zhao}}]{panNonlinearSteeringCriteria2021}%
  \BibitemOpen
  \bibfield  {author} {\bibinfo {author} {\bibfnamefont {G.-Z.}\ \bibnamefont
  {Pan}}, \bibinfo {author} {\bibfnamefont {M.}~\bibnamefont {Yang}}, \bibinfo
  {author} {\bibfnamefont {H.}~\bibnamefont {Yuan}}, \bibinfo {author}
  {\bibfnamefont {G.}~\bibnamefont {Zhang}}, \ and\ \bibinfo {author}
  {\bibfnamefont {J.-L.}\ \bibnamefont {Zhao}},\ }\href {\doibase
  10.1007/s11128-020-02954-5} {\bibfield  {journal} {\bibinfo  {journal}
  {Quantum Information Processing}\ }\textbf {\bibinfo {volume} {20}},\
  \bibinfo {pages} {48} (\bibinfo {year} {2021})}\BibitemShut {NoStop}%
\bibitem [{\citenamefont {Saunders}\ \emph {et~al.}(2010)\citenamefont
  {Saunders}, \citenamefont {Jones}, \citenamefont {Wiseman},\ and\
  \citenamefont {Pryde}}]{SaundersExperimental-EPR-steering-using2010}%
  \BibitemOpen
  \bibfield  {author} {\bibinfo {author} {\bibfnamefont {D.~J.}\ \bibnamefont
  {Saunders}}, \bibinfo {author} {\bibfnamefont {S.~J.}\ \bibnamefont {Jones}},
  \bibinfo {author} {\bibfnamefont {H.~M.}\ \bibnamefont {Wiseman}}, \ and\
  \bibinfo {author} {\bibfnamefont {G.~J.}\ \bibnamefont {Pryde}},\ }\href
  {\doibase 10.1038/nphys1766} {\bibfield  {journal} {\bibinfo  {journal}
  {Nature Physics}\ }\textbf {\bibinfo {volume} {6}},\ \bibinfo {pages} {845}
  (\bibinfo {year} {2010})}\BibitemShut {NoStop}%
\bibitem [{\citenamefont {Leach}\ \emph {et~al.}(2010)\citenamefont {Leach},
  \citenamefont {Jack}, \citenamefont {Romero}, \citenamefont {Jha},
  \citenamefont {Yao}, \citenamefont {Franke-Arnold}, \citenamefont {Ireland},
  \citenamefont {Boyd}, \citenamefont {Barnett},\ and\ \citenamefont
  {Padgett}}]{LeachQuantum-Correlations-in-Optical2010}%
  \BibitemOpen
  \bibfield  {author} {\bibinfo {author} {\bibfnamefont {J.}~\bibnamefont
  {Leach}}, \bibinfo {author} {\bibfnamefont {B.}~\bibnamefont {Jack}},
  \bibinfo {author} {\bibfnamefont {J.}~\bibnamefont {Romero}}, \bibinfo
  {author} {\bibfnamefont {A.~K.}\ \bibnamefont {Jha}}, \bibinfo {author}
  {\bibfnamefont {A.~M.}\ \bibnamefont {Yao}}, \bibinfo {author} {\bibfnamefont
  {S.}~\bibnamefont {Franke-Arnold}}, \bibinfo {author} {\bibfnamefont {D.~G.}\
  \bibnamefont {Ireland}}, \bibinfo {author} {\bibfnamefont {R.~W.}\
  \bibnamefont {Boyd}}, \bibinfo {author} {\bibfnamefont {S.~M.}\ \bibnamefont
  {Barnett}}, \ and\ \bibinfo {author} {\bibfnamefont {M.~J.}\ \bibnamefont
  {Padgett}},\ }\href {\doibase 10.1126/science.1190523} {\bibfield  {journal}
  {\bibinfo  {journal} {Science}\ }\textbf {\bibinfo {volume} {329}},\ \bibinfo
  {pages} {662} (\bibinfo {year} {2010})}\BibitemShut {NoStop}%
\bibitem [{\citenamefont {Walborn}\ \emph {et~al.}(2011)\citenamefont
  {Walborn}, \citenamefont {Salles}, \citenamefont {Gomes}, \citenamefont
  {Toscano},\ and\ \citenamefont
  {Souto~Ribeiro}}]{WalbornRevealing-Hidden-Einstein-Podolsky-Rosen2011}%
  \BibitemOpen
  \bibfield  {author} {\bibinfo {author} {\bibfnamefont {S.~P.}\ \bibnamefont
  {Walborn}}, \bibinfo {author} {\bibfnamefont {A.}~\bibnamefont {Salles}},
  \bibinfo {author} {\bibfnamefont {R.~M.}\ \bibnamefont {Gomes}}, \bibinfo
  {author} {\bibfnamefont {F.}~\bibnamefont {Toscano}}, \ and\ \bibinfo
  {author} {\bibfnamefont {P.~H.}\ \bibnamefont {Souto~Ribeiro}},\ }\href
  {\doibase 10.1103/PhysRevLett.106.130402} {\bibfield  {journal} {\bibinfo
  {journal} {Phys. Rev. Lett.}\ }\textbf {\bibinfo {volume} {106}},\ \bibinfo
  {pages} {130402} (\bibinfo {year} {2011})}\BibitemShut {NoStop}%
\bibitem [{\citenamefont {Wittmann}\ \emph {et~al.}(2012)\citenamefont
  {Wittmann}, \citenamefont {Ramelow}, \citenamefont {Steinlechner},
  \citenamefont {Langford}, \citenamefont {Brunner}, \citenamefont {Wiseman},
  \citenamefont {Ursin},\ and\ \citenamefont
  {Zeilinger}}]{wittmannLoopholefreeEinsteinPodolsky2012}%
  \BibitemOpen
  \bibfield  {author} {\bibinfo {author} {\bibfnamefont {B.}~\bibnamefont
  {Wittmann}}, \bibinfo {author} {\bibfnamefont {S.}~\bibnamefont {Ramelow}},
  \bibinfo {author} {\bibfnamefont {F.}~\bibnamefont {Steinlechner}}, \bibinfo
  {author} {\bibfnamefont {N.~K.}\ \bibnamefont {Langford}}, \bibinfo {author}
  {\bibfnamefont {N.}~\bibnamefont {Brunner}}, \bibinfo {author} {\bibfnamefont
  {H.~M.}\ \bibnamefont {Wiseman}}, \bibinfo {author} {\bibfnamefont
  {R.}~\bibnamefont {Ursin}}, \ and\ \bibinfo {author} {\bibfnamefont
  {A.}~\bibnamefont {Zeilinger}},\ }\href {\doibase
  10.1088/1367-2630/14/5/053030} {\bibfield  {journal} {\bibinfo  {journal}
  {New Journal of Physics}\ }\textbf {\bibinfo {volume} {14}},\ \bibinfo
  {pages} {053030} (\bibinfo {year} {2012})}\BibitemShut {NoStop}%
\bibitem [{\citenamefont {Nery}\ \emph {et~al.}(2020)\citenamefont {Nery},
  \citenamefont {Taddei}, \citenamefont {Sahium}, \citenamefont {Walborn},
  \citenamefont {Aolita},\ and\ \citenamefont
  {Aguilar}}]{neryDistillationQuantumSteering2020}%
  \BibitemOpen
  \bibfield  {author} {\bibinfo {author} {\bibfnamefont {R.~V.}\ \bibnamefont
  {Nery}}, \bibinfo {author} {\bibfnamefont {M.~M.}\ \bibnamefont {Taddei}},
  \bibinfo {author} {\bibfnamefont {P.}~\bibnamefont {Sahium}}, \bibinfo
  {author} {\bibfnamefont {S.~P.}\ \bibnamefont {Walborn}}, \bibinfo {author}
  {\bibfnamefont {L.}~\bibnamefont {Aolita}}, \ and\ \bibinfo {author}
  {\bibfnamefont {G.~H.}\ \bibnamefont {Aguilar}},\ }\href {\doibase
  10.1103/PhysRevLett.124.120402} {\bibfield  {journal} {\bibinfo  {journal}
  {Physical Review Letters}\ }\textbf {\bibinfo {volume} {124}},\ \bibinfo
  {pages} {120402} (\bibinfo {year} {2020})}\BibitemShut {NoStop}%
\bibitem [{\citenamefont {Zhao}\ \emph {et~al.}(2020)\citenamefont {Zhao},
  \citenamefont {Ku}, \citenamefont {Chen}, \citenamefont {Chen}, \citenamefont
  {Nori}, \citenamefont {Xiang}, \citenamefont {Li}, \citenamefont {Guo},\ and\
  \citenamefont
  {Chen}}]{zhaoExperimentalDemonstrationMeasurementdeviceindependent2020}%
  \BibitemOpen
  \bibfield  {author} {\bibinfo {author} {\bibfnamefont {Y.-Y.}\ \bibnamefont
  {Zhao}}, \bibinfo {author} {\bibfnamefont {H.-Y.}\ \bibnamefont {Ku}},
  \bibinfo {author} {\bibfnamefont {S.-L.}\ \bibnamefont {Chen}}, \bibinfo
  {author} {\bibfnamefont {H.-B.}\ \bibnamefont {Chen}}, \bibinfo {author}
  {\bibfnamefont {F.}~\bibnamefont {Nori}}, \bibinfo {author} {\bibfnamefont
  {G.-Y.}\ \bibnamefont {Xiang}}, \bibinfo {author} {\bibfnamefont {C.-F.}\
  \bibnamefont {Li}}, \bibinfo {author} {\bibfnamefont {G.-C.}\ \bibnamefont
  {Guo}}, \ and\ \bibinfo {author} {\bibfnamefont {Y.-N.}\ \bibnamefont
  {Chen}},\ }\href {\doibase 10.1038/s41534-020-00307-9} {\bibfield  {journal}
  {\bibinfo  {journal} {npj Quantum Information}\ }\textbf {\bibinfo {volume}
  {6}},\ \bibinfo {pages} {1} (\bibinfo {year} {2020})}\BibitemShut {NoStop}%
\bibitem [{\citenamefont {Wollmann}\ \emph {et~al.}(2020)\citenamefont
  {Wollmann}, \citenamefont {Uola},\ and\ \citenamefont
  {Costa}}]{wollmannExperimentalDemonstrationRobust2020}%
  \BibitemOpen
  \bibfield  {author} {\bibinfo {author} {\bibfnamefont {S.}~\bibnamefont
  {Wollmann}}, \bibinfo {author} {\bibfnamefont {R.}~\bibnamefont {Uola}}, \
  and\ \bibinfo {author} {\bibfnamefont {A.~C.~S.}\ \bibnamefont {Costa}},\
  }\href {\doibase 10.1103/PhysRevLett.125.020404} {\bibfield  {journal}
  {\bibinfo  {journal} {Physical Review Letters}\ }\textbf {\bibinfo {volume}
  {125}},\ \bibinfo {pages} {020404} (\bibinfo {year} {2020})}\BibitemShut
  {NoStop}%
\bibitem [{\citenamefont {Wiseman}\ and\ \citenamefont
  {Gambetta}(2012)}]{WisemanAre-Dynamical-Quantum-Jumps2012}%
  \BibitemOpen
  \bibfield  {author} {\bibinfo {author} {\bibfnamefont {H.~M.}\ \bibnamefont
  {Wiseman}}\ and\ \bibinfo {author} {\bibfnamefont {J.~M.}\ \bibnamefont
  {Gambetta}},\ }\href {\doibase 10.1103/PhysRevLett.108.220402} {\bibfield
  {journal} {\bibinfo  {journal} {Phys. Rev. Lett.}\ }\textbf {\bibinfo
  {volume} {108}},\ \bibinfo {pages} {220402} (\bibinfo {year}
  {2012})}\BibitemShut {NoStop}%
\bibitem [{\citenamefont {Beyer}\ \emph {et~al.}(2018)\citenamefont {Beyer},
  \citenamefont {Luoma},\ and\ \citenamefont
  {Strunz}}]{BeyerCollision-model-approach-to-steering2018}%
  \BibitemOpen
  \bibfield  {author} {\bibinfo {author} {\bibfnamefont {K.}~\bibnamefont
  {Beyer}}, \bibinfo {author} {\bibfnamefont {K.}~\bibnamefont {Luoma}}, \ and\
  \bibinfo {author} {\bibfnamefont {W.~T.}\ \bibnamefont {Strunz}},\ }\href
  {\doibase 10.1103/PhysRevA.97.032113} {\bibfield  {journal} {\bibinfo
  {journal} {Phys. Rev. A}\ }\textbf {\bibinfo {volume} {97}},\ \bibinfo
  {pages} {032113} (\bibinfo {year} {2018})}\BibitemShut {NoStop}%
\bibitem [{\citenamefont {Feynman}(1982)}]{FeynmanSimulating-physics-with1982}%
  \BibitemOpen
  \bibfield  {author} {\bibinfo {author} {\bibfnamefont {R.~P.}\ \bibnamefont
  {Feynman}},\ }\href {\doibase 10.1007/BF02650179} {\bibfield  {journal}
  {\bibinfo  {journal} {International Journal of Theoretical Physics}\ }\textbf
  {\bibinfo {volume} {21}},\ \bibinfo {pages} {467} (\bibinfo {year}
  {1982})}\BibitemShut {NoStop}%
\bibitem [{ibm(2021)}]{ibmq2021}%
  \BibitemOpen
  \href@noop {} {\enquote {\bibinfo {title} {\mbox{IBM Q}uantum},}\ }\bibinfo
  {howpublished} {https://quantum-computing.ibm.com/} (\bibinfo {year}
  {2021})\BibitemShut {NoStop}%
\bibitem [{\citenamefont {Wang}\ \emph {et~al.}(2018)\citenamefont {Wang},
  \citenamefont {Li}, \citenamefont {Yin},\ and\ \citenamefont
  {Zeng}}]{wang16qubitIBMUniversal2018}%
  \BibitemOpen
  \bibfield  {author} {\bibinfo {author} {\bibfnamefont {Y.}~\bibnamefont
  {Wang}}, \bibinfo {author} {\bibfnamefont {Y.}~\bibnamefont {Li}}, \bibinfo
  {author} {\bibfnamefont {Z.-q.}\ \bibnamefont {Yin}}, \ and\ \bibinfo
  {author} {\bibfnamefont {B.}~\bibnamefont {Zeng}},\ }\href {\doibase
  10.1038/s41534-018-0095-x} {\bibfield  {journal} {\bibinfo  {journal} {npj
  Quantum Information}\ }\textbf {\bibinfo {volume} {4}},\ \bibinfo {pages} {1}
  (\bibinfo {year} {2018})}\BibitemShut {NoStop}%
\bibitem [{\citenamefont {Koppenh{\"o}fer}\ \emph {et~al.}(2020)\citenamefont
  {Koppenh{\"o}fer}, \citenamefont {Bruder},\ and\ \citenamefont
  {Roulet}}]{koppenhoferQuantumSynchronizationIBM2020}%
  \BibitemOpen
  \bibfield  {author} {\bibinfo {author} {\bibfnamefont {M.}~\bibnamefont
  {Koppenh{\"o}fer}}, \bibinfo {author} {\bibfnamefont {C.}~\bibnamefont
  {Bruder}}, \ and\ \bibinfo {author} {\bibfnamefont {A.}~\bibnamefont
  {Roulet}},\ }\href {\doibase 10.1103/PhysRevResearch.2.023026} {\bibfield
  {journal} {\bibinfo  {journal} {Physical Review Research}\ }\textbf {\bibinfo
  {volume} {2}},\ \bibinfo {pages} {023026} (\bibinfo {year}
  {2020})}\BibitemShut {NoStop}%
\bibitem [{\citenamefont {Riedel~G{\aa}rding}\ \emph
  {et~al.}(2021)\citenamefont {Riedel~G{\aa}rding}, \citenamefont {Schwaller},
  \citenamefont {Chan}, \citenamefont {Chang}, \citenamefont {Bosch},
  \citenamefont {Gessler}, \citenamefont {Laborde}, \citenamefont {Hernandez},
  \citenamefont {Si}, \citenamefont {Dupertuis},\ and\ \citenamefont
  {Macris}}]{riedelgardingBellDiagonalWerner2021}%
  \BibitemOpen
  \bibfield  {author} {\bibinfo {author} {\bibfnamefont {E.}~\bibnamefont
  {Riedel~G{\aa}rding}}, \bibinfo {author} {\bibfnamefont {N.}~\bibnamefont
  {Schwaller}}, \bibinfo {author} {\bibfnamefont {C.~L.}\ \bibnamefont {Chan}},
  \bibinfo {author} {\bibfnamefont {S.~Y.}\ \bibnamefont {Chang}}, \bibinfo
  {author} {\bibfnamefont {S.}~\bibnamefont {Bosch}}, \bibinfo {author}
  {\bibfnamefont {F.}~\bibnamefont {Gessler}}, \bibinfo {author} {\bibfnamefont
  {W.~R.}\ \bibnamefont {Laborde}}, \bibinfo {author} {\bibfnamefont {J.~N.}\
  \bibnamefont {Hernandez}}, \bibinfo {author} {\bibfnamefont {X.}~\bibnamefont
  {Si}}, \bibinfo {author} {\bibfnamefont {M.-A.}\ \bibnamefont {Dupertuis}}, \
  and\ \bibinfo {author} {\bibfnamefont {N.}~\bibnamefont {Macris}},\ }\href
  {\doibase 10.3390/e23070797} {\bibfield  {journal} {\bibinfo  {journal}
  {Entropy}\ }\textbf {\bibinfo {volume} {23}},\ \bibinfo {pages} {797}
  (\bibinfo {year} {2021})}\BibitemShut {NoStop}%
\bibitem [{\citenamefont {Leontica}\ \emph {et~al.}(2021)\citenamefont
  {Leontica}, \citenamefont {Tennie},\ and\ \citenamefont
  {Farrow}}]{leonticaSimulatingMoleculesCloudbased2021}%
  \BibitemOpen
  \bibfield  {author} {\bibinfo {author} {\bibfnamefont {S.}~\bibnamefont
  {Leontica}}, \bibinfo {author} {\bibfnamefont {F.}~\bibnamefont {Tennie}}, \
  and\ \bibinfo {author} {\bibfnamefont {T.}~\bibnamefont {Farrow}},\ }\href
  {\doibase 10.1038/s42005-021-00616-1} {\bibfield  {journal} {\bibinfo
  {journal} {Communications Physics}\ }\textbf {\bibinfo {volume} {4}},\
  \bibinfo {pages} {1} (\bibinfo {year} {2021})}\BibitemShut {NoStop}%
\bibitem [{\citenamefont {Choo}\ \emph {et~al.}(2018)\citenamefont {Choo},
  \citenamefont {{von Keyserlingk}}, \citenamefont {Regnault},\ and\
  \citenamefont {Neupert}}]{chooMeasurementEntanglementSpectrum2018}%
  \BibitemOpen
  \bibfield  {author} {\bibinfo {author} {\bibfnamefont {K.}~\bibnamefont
  {Choo}}, \bibinfo {author} {\bibfnamefont {C.~W.}\ \bibnamefont {{von
  Keyserlingk}}}, \bibinfo {author} {\bibfnamefont {N.}~\bibnamefont
  {Regnault}}, \ and\ \bibinfo {author} {\bibfnamefont {T.}~\bibnamefont
  {Neupert}},\ }\href {\doibase 10.1103/PhysRevLett.121.086808} {\bibfield
  {journal} {\bibinfo  {journal} {Physical Review Letters}\ }\textbf {\bibinfo
  {volume} {121}},\ \bibinfo {pages} {086808} (\bibinfo {year}
  {2018})}\BibitemShut {NoStop}%
\bibitem [{\citenamefont {{Garc{\'\i}a-P{\'e}rez}}\ \emph
  {et~al.}(2020)\citenamefont {{Garc{\'\i}a-P{\'e}rez}}, \citenamefont
  {Rossi},\ and\ \citenamefont
  {Maniscalco}}]{garcia-perezIBMExperienceVersatile2020}%
  \BibitemOpen
  \bibfield  {author} {\bibinfo {author} {\bibfnamefont {G.}~\bibnamefont
  {{Garc{\'\i}a-P{\'e}rez}}}, \bibinfo {author} {\bibfnamefont {M.~A.~C.}\
  \bibnamefont {Rossi}}, \ and\ \bibinfo {author} {\bibfnamefont
  {S.}~\bibnamefont {Maniscalco}},\ }\href {\doibase 10.1038/s41534-019-0235-y}
  {\bibfield  {journal} {\bibinfo  {journal} {npj Quantum Information}\
  }\textbf {\bibinfo {volume} {6}},\ \bibinfo {pages} {1} (\bibinfo {year}
  {2020})}\BibitemShut {NoStop}%
\bibitem [{\citenamefont {K{\"u}mmerer}\ and\ \citenamefont
  {Maassen}(1987)}]{kummererEssentiallyCommutativeDilations1987}%
  \BibitemOpen
  \bibfield  {author} {\bibinfo {author} {\bibfnamefont {B.}~\bibnamefont
  {K{\"u}mmerer}}\ and\ \bibinfo {author} {\bibfnamefont {H.}~\bibnamefont
  {Maassen}},\ }\href {\doibase 10.1007/BF01205670} {\bibfield  {journal}
  {\bibinfo  {journal} {Communications in Mathematical Physics}\ }\textbf
  {\bibinfo {volume} {109}},\ \bibinfo {pages} {1} (\bibinfo {year}
  {1987})}\BibitemShut {NoStop}%
\bibitem [{\citenamefont {Landau}\ and\ \citenamefont
  {Streater}(1993)}]{landauBirkhoffTheoremDoubly1993}%
  \BibitemOpen
  \bibfield  {author} {\bibinfo {author} {\bibfnamefont {L.~J.}\ \bibnamefont
  {Landau}}\ and\ \bibinfo {author} {\bibfnamefont {R.~F.}\ \bibnamefont
  {Streater}},\ }\href {\doibase 10.1016/0024-3795(93)90274-R} {\bibfield
  {journal} {\bibinfo  {journal} {Linear Algebra and its Applications}\
  }\textbf {\bibinfo {volume} {193}},\ \bibinfo {pages} {107} (\bibinfo {year}
  {1993})}\BibitemShut {NoStop}%
\bibitem [{\citenamefont {Buscemi}\ \emph {et~al.}(2005)\citenamefont
  {Buscemi}, \citenamefont {Chiribella},\ and\ \citenamefont
  {Mauro~D'Ariano}}]{buscemiInvertingQuantumDecoherence2005}%
  \BibitemOpen
  \bibfield  {author} {\bibinfo {author} {\bibfnamefont {F.}~\bibnamefont
  {Buscemi}}, \bibinfo {author} {\bibfnamefont {G.}~\bibnamefont {Chiribella}},
  \ and\ \bibinfo {author} {\bibfnamefont {G.}~\bibnamefont {Mauro~D'Ariano}},\
  }\href {\doibase 10.1103/PhysRevLett.95.090501} {\bibfield  {journal}
  {\bibinfo  {journal} {Physical Review Letters}\ }\textbf {\bibinfo {volume}
  {95}},\ \bibinfo {pages} {090501} (\bibinfo {year} {2005})}\BibitemShut
  {NoStop}%
\bibitem [{\citenamefont {Pernice}\ \emph {et~al.}(2012)\citenamefont
  {Pernice}, \citenamefont {Helm},\ and\ \citenamefont
  {Strunz}}]{perniceSystemEnvironmentCorrelations2012}%
  \BibitemOpen
  \bibfield  {author} {\bibinfo {author} {\bibfnamefont {A.}~\bibnamefont
  {Pernice}}, \bibinfo {author} {\bibfnamefont {J.}~\bibnamefont {Helm}}, \
  and\ \bibinfo {author} {\bibfnamefont {W.~T.}\ \bibnamefont {Strunz}},\
  }\href {\doibase 10.1088/0953-4075/45/15/154005} {\bibfield  {journal}
  {\bibinfo  {journal} {Journal of Physics B: Atomic, Molecular and Optical
  Physics}\ }\textbf {\bibinfo {volume} {45}},\ \bibinfo {pages} {154005}
  (\bibinfo {year} {2012})}\BibitemShut {NoStop}%
\bibitem [{\citenamefont {Vandenberghe}\ and\ \citenamefont
  {Boyd}(1996)}]{VandenbergheSemidefinite-Programming1996}%
  \BibitemOpen
  \bibfield  {author} {\bibinfo {author} {\bibfnamefont {L.}~\bibnamefont
  {Vandenberghe}}\ and\ \bibinfo {author} {\bibfnamefont {S.}~\bibnamefont
  {Boyd}},\ }\bibfield  {booktitle} {\emph {\bibinfo {booktitle} {SIAM
  Review}},\ }\href {\doibase 10.1137/1038003} {\bibfield  {journal} {\bibinfo
  {journal} {SIAM Review}\ }\textbf {\bibinfo {volume} {38}},\ \bibinfo {pages}
  {49} (\bibinfo {year} {1996})}\BibitemShut {NoStop}%
\bibitem [{Note1()}]{Note1}%
  \BibitemOpen
  \bibinfo {note} {Without the noise the optimization would fail since the
  steering weight would already be saturated by the first two pure
  ensembles.}\BibitemShut {Stop}%
\bibitem [{\citenamefont {Gu{\c{t}}{\u{a}}}\ \emph {et~al.}(2020)\citenamefont
  {Gu{\c{t}}{\u{a}}}, \citenamefont {Kahn}, \citenamefont {Kueng},\ and\
  \citenamefont {Tropp}}]{GutaFast-state-tomography2020}%
  \BibitemOpen
  \bibfield  {author} {\bibinfo {author} {\bibfnamefont {M.}~\bibnamefont
  {Gu{\c{t}}{\u{a}}}}, \bibinfo {author} {\bibfnamefont {J.}~\bibnamefont
  {Kahn}}, \bibinfo {author} {\bibfnamefont {R.}~\bibnamefont {Kueng}}, \ and\
  \bibinfo {author} {\bibfnamefont {J.~A.}\ \bibnamefont {Tropp}},\ }\href
  {\doibase 10.1088/1751-8121/ab8111} {\bibfield  {journal} {\bibinfo
  {journal} {Journal of Physics A: Mathematical and Theoretical}\ }\textbf
  {\bibinfo {volume} {53}},\ \bibinfo {pages} {204001} (\bibinfo {year}
  {2020})}\BibitemShut {NoStop}%
\bibitem [{\citenamefont {D'Ariano}\ and\ \citenamefont
  {Perinotti}(2007)}]{DArianoOptimal-Data-Processing2007}%
  \BibitemOpen
  \bibfield  {author} {\bibinfo {author} {\bibfnamefont {G.~M.}\ \bibnamefont
  {D'Ariano}}\ and\ \bibinfo {author} {\bibfnamefont {P.}~\bibnamefont
  {Perinotti}},\ }\href {\doibase 10.1103/PhysRevLett.98.020403} {\bibfield
  {journal} {\bibinfo  {journal} {Phys. Rev. Lett.}\ }\textbf {\bibinfo
  {volume} {98}},\ \bibinfo {pages} {020403} (\bibinfo {year}
  {2007})}\BibitemShut {NoStop}%
\bibitem [{\citenamefont {Sarovar}\ \emph {et~al.}(2020)\citenamefont
  {Sarovar}, \citenamefont {Proctor}, \citenamefont {Rudinger}, \citenamefont
  {Young}, \citenamefont {Nielsen},\ and\ \citenamefont
  {{Blume-Kohout}}}]{SarovarDetectingcrosstalkerrors2020}%
  \BibitemOpen
  \bibfield  {author} {\bibinfo {author} {\bibfnamefont {M.}~\bibnamefont
  {Sarovar}}, \bibinfo {author} {\bibfnamefont {T.}~\bibnamefont {Proctor}},
  \bibinfo {author} {\bibfnamefont {K.}~\bibnamefont {Rudinger}}, \bibinfo
  {author} {\bibfnamefont {K.}~\bibnamefont {Young}}, \bibinfo {author}
  {\bibfnamefont {E.}~\bibnamefont {Nielsen}}, \ and\ \bibinfo {author}
  {\bibfnamefont {R.}~\bibnamefont {{Blume-Kohout}}},\ }\href {\doibase
  10.22331/q-2020-09-11-321} {\bibfield  {journal} {\bibinfo  {journal}
  {Quantum}\ }\textbf {\bibinfo {volume} {4}},\ \bibinfo {pages} {321}
  (\bibinfo {year} {2020})}\BibitemShut {NoStop}%
\bibitem [{\citenamefont {Hradil}(1997)}]{hradilQuantumstateEstimation1997}%
  \BibitemOpen
  \bibfield  {author} {\bibinfo {author} {\bibfnamefont {Z.}~\bibnamefont
  {Hradil}},\ }\href {\doibase 10.1103/PhysRevA.55.R1561} {\bibfield  {journal}
  {\bibinfo  {journal} {Physical Review A}\ }\textbf {\bibinfo {volume} {55}},\
  \bibinfo {pages} {R1561} (\bibinfo {year} {1997})}\BibitemShut {NoStop}%
\bibitem [{\citenamefont
  {{Blume-Kohout}}(2010)}]{blume-kohoutHedgedMaximumLikelihood2010}%
  \BibitemOpen
  \bibfield  {author} {\bibinfo {author} {\bibfnamefont {R.}~\bibnamefont
  {{Blume-Kohout}}},\ }\href {\doibase 10.1103/PhysRevLett.105.200504}
  {\bibfield  {journal} {\bibinfo  {journal} {Physical Review Letters}\
  }\textbf {\bibinfo {volume} {105}},\ \bibinfo {pages} {200504} (\bibinfo
  {year} {2010})}\BibitemShut {NoStop}%
\bibitem [{\citenamefont {Smolin}\ \emph {et~al.}(2012)\citenamefont {Smolin},
  \citenamefont {Gambetta},\ and\ \citenamefont
  {Smith}}]{smolinEfficientMethodComputing2012}%
  \BibitemOpen
  \bibfield  {author} {\bibinfo {author} {\bibfnamefont {J.~A.}\ \bibnamefont
  {Smolin}}, \bibinfo {author} {\bibfnamefont {J.~M.}\ \bibnamefont
  {Gambetta}}, \ and\ \bibinfo {author} {\bibfnamefont {G.}~\bibnamefont
  {Smith}},\ }\href {\doibase 10.1103/PhysRevLett.108.070502} {\bibfield
  {journal} {\bibinfo  {journal} {Physical Review Letters}\ }\textbf {\bibinfo
  {volume} {108}},\ \bibinfo {pages} {070502} (\bibinfo {year}
  {2012})}\BibitemShut {NoStop}%
\bibitem [{\citenamefont
  {Blume-Kohout}(2010)}]{Blume-KohoutOptimal-reliable-estimation2010}%
  \BibitemOpen
  \bibfield  {author} {\bibinfo {author} {\bibfnamefont {R.}~\bibnamefont
  {Blume-Kohout}},\ }\href {\doibase 10.1088/1367-2630/12/4/043034} {\bibfield
  {journal} {\bibinfo  {journal} {New Journal of Physics}\ }\textbf {\bibinfo
  {volume} {12}},\ \bibinfo {pages} {043034} (\bibinfo {year}
  {2010})}\BibitemShut {NoStop}%
\bibitem [{\citenamefont {Husz{\'a}r}\ and\ \citenamefont
  {Houlsby}(2012)}]{huszarAdaptiveBayesianQuantum2012}%
  \BibitemOpen
  \bibfield  {author} {\bibinfo {author} {\bibfnamefont {F.}~\bibnamefont
  {Husz{\'a}r}}\ and\ \bibinfo {author} {\bibfnamefont {N.~M.~T.}\ \bibnamefont
  {Houlsby}},\ }\href {\doibase 10.1103/PhysRevA.85.052120} {\bibfield
  {journal} {\bibinfo  {journal} {Physical Review A}\ }\textbf {\bibinfo
  {volume} {85}},\ \bibinfo {pages} {052120} (\bibinfo {year}
  {2012})}\BibitemShut {NoStop}%
\bibitem [{\citenamefont {Lukens}\ \emph {et~al.}(2020)\citenamefont {Lukens},
  \citenamefont {Law}, \citenamefont {Jasra},\ and\ \citenamefont
  {Lougovski}}]{lukensPracticalEfficientApproach2020}%
  \BibitemOpen
  \bibfield  {author} {\bibinfo {author} {\bibfnamefont {J.~M.}\ \bibnamefont
  {Lukens}}, \bibinfo {author} {\bibfnamefont {K.~J.~H.}\ \bibnamefont {Law}},
  \bibinfo {author} {\bibfnamefont {A.}~\bibnamefont {Jasra}}, \ and\ \bibinfo
  {author} {\bibfnamefont {P.}~\bibnamefont {Lougovski}},\ }\href {\doibase
  10.1088/1367-2630/ab8efa} {\bibfield  {journal} {\bibinfo  {journal} {New
  Journal of Physics}\ }\textbf {\bibinfo {volume} {22}},\ \bibinfo {pages}
  {063038} (\bibinfo {year} {2020})}\BibitemShut {NoStop}%
\bibitem [{\citenamefont {Opatrn{\'y}}\ \emph {et~al.}(1997)\citenamefont
  {Opatrn{\'y}}, \citenamefont {Welsch},\ and\ \citenamefont
  {Vogel}}]{opatrnyLeastsquaresInversionDensitymatrix1997}%
  \BibitemOpen
  \bibfield  {author} {\bibinfo {author} {\bibfnamefont {T.}~\bibnamefont
  {Opatrn{\'y}}}, \bibinfo {author} {\bibfnamefont {D.-G.}\ \bibnamefont
  {Welsch}}, \ and\ \bibinfo {author} {\bibfnamefont {W.}~\bibnamefont
  {Vogel}},\ }\href {\doibase 10.1103/PhysRevA.56.1788} {\bibfield  {journal}
  {\bibinfo  {journal} {Physical Review A}\ }\textbf {\bibinfo {volume} {56}},\
  \bibinfo {pages} {1788} (\bibinfo {year} {1997})}\BibitemShut {NoStop}%
\bibitem [{\citenamefont {James}\ \emph {et~al.}(2001)\citenamefont {James},
  \citenamefont {Kwiat}, \citenamefont {Munro},\ and\ \citenamefont
  {White}}]{jamesMeasurementQubits2001}%
  \BibitemOpen
  \bibfield  {author} {\bibinfo {author} {\bibfnamefont {D.~F.~V.}\
  \bibnamefont {James}}, \bibinfo {author} {\bibfnamefont {P.~G.}\ \bibnamefont
  {Kwiat}}, \bibinfo {author} {\bibfnamefont {W.~J.}\ \bibnamefont {Munro}}, \
  and\ \bibinfo {author} {\bibfnamefont {A.~G.}\ \bibnamefont {White}},\ }\href
  {\doibase 10.1103/PhysRevA.64.052312} {\bibfield  {journal} {\bibinfo
  {journal} {Physical Review A}\ }\textbf {\bibinfo {volume} {64}},\ \bibinfo
  {pages} {052312} (\bibinfo {year} {2001})}\BibitemShut {NoStop}%
\bibitem [{\citenamefont {Qi}\ \emph {et~al.}(2013)\citenamefont {Qi},
  \citenamefont {Hou}, \citenamefont {Li}, \citenamefont {Dong}, \citenamefont
  {Xiang},\ and\ \citenamefont {Guo}}]{qiQuantumStateTomography2013}%
  \BibitemOpen
  \bibfield  {author} {\bibinfo {author} {\bibfnamefont {B.}~\bibnamefont
  {Qi}}, \bibinfo {author} {\bibfnamefont {Z.}~\bibnamefont {Hou}}, \bibinfo
  {author} {\bibfnamefont {L.}~\bibnamefont {Li}}, \bibinfo {author}
  {\bibfnamefont {D.}~\bibnamefont {Dong}}, \bibinfo {author} {\bibfnamefont
  {G.}~\bibnamefont {Xiang}}, \ and\ \bibinfo {author} {\bibfnamefont
  {G.}~\bibnamefont {Guo}},\ }\href {\doibase 10.1038/srep03496} {\bibfield
  {journal} {\bibinfo  {journal} {Scientific Reports}\ }\textbf {\bibinfo
  {volume} {3}},\ \bibinfo {pages} {3496} (\bibinfo {year} {2013})}\BibitemShut
  {NoStop}%
\bibitem [{\citenamefont {Hou}\ \emph {et~al.}(2016)\citenamefont {Hou},
  \citenamefont {Zhong}, \citenamefont {Tian}, \citenamefont {Dong},
  \citenamefont {Qi}, \citenamefont {Li}, \citenamefont {Wang}, \citenamefont
  {Nori}, \citenamefont {Xiang}, \citenamefont {Li},\ and\ \citenamefont
  {Guo}}]{houFullReconstruction14qubit2016}%
  \BibitemOpen
  \bibfield  {author} {\bibinfo {author} {\bibfnamefont {Z.}~\bibnamefont
  {Hou}}, \bibinfo {author} {\bibfnamefont {H.-S.}\ \bibnamefont {Zhong}},
  \bibinfo {author} {\bibfnamefont {Y.}~\bibnamefont {Tian}}, \bibinfo {author}
  {\bibfnamefont {D.}~\bibnamefont {Dong}}, \bibinfo {author} {\bibfnamefont
  {B.}~\bibnamefont {Qi}}, \bibinfo {author} {\bibfnamefont {L.}~\bibnamefont
  {Li}}, \bibinfo {author} {\bibfnamefont {Y.}~\bibnamefont {Wang}}, \bibinfo
  {author} {\bibfnamefont {F.}~\bibnamefont {Nori}}, \bibinfo {author}
  {\bibfnamefont {G.-Y.}\ \bibnamefont {Xiang}}, \bibinfo {author}
  {\bibfnamefont {C.-F.}\ \bibnamefont {Li}}, \ and\ \bibinfo {author}
  {\bibfnamefont {G.-C.}\ \bibnamefont {Guo}},\ }\href {\doibase
  10.1088/1367-2630/18/8/083036} {\bibfield  {journal} {\bibinfo  {journal}
  {New Journal of Physics}\ }\textbf {\bibinfo {volume} {18}},\ \bibinfo
  {pages} {083036} (\bibinfo {year} {2016})}\BibitemShut {NoStop}%
\bibitem [{\citenamefont {Qi}\ \emph {et~al.}(2017)\citenamefont {Qi},
  \citenamefont {Hou}, \citenamefont {Wang}, \citenamefont {Dong},
  \citenamefont {Zhong}, \citenamefont {Li}, \citenamefont {Xiang},
  \citenamefont {Wiseman}, \citenamefont {Li},\ and\ \citenamefont
  {Guo}}]{qiAdaptiveQuantumState2017}%
  \BibitemOpen
  \bibfield  {author} {\bibinfo {author} {\bibfnamefont {B.}~\bibnamefont
  {Qi}}, \bibinfo {author} {\bibfnamefont {Z.}~\bibnamefont {Hou}}, \bibinfo
  {author} {\bibfnamefont {Y.}~\bibnamefont {Wang}}, \bibinfo {author}
  {\bibfnamefont {D.}~\bibnamefont {Dong}}, \bibinfo {author} {\bibfnamefont
  {H.-S.}\ \bibnamefont {Zhong}}, \bibinfo {author} {\bibfnamefont
  {L.}~\bibnamefont {Li}}, \bibinfo {author} {\bibfnamefont {G.-Y.}\
  \bibnamefont {Xiang}}, \bibinfo {author} {\bibfnamefont {H.~M.}\ \bibnamefont
  {Wiseman}}, \bibinfo {author} {\bibfnamefont {C.-F.}\ \bibnamefont {Li}}, \
  and\ \bibinfo {author} {\bibfnamefont {G.-C.}\ \bibnamefont {Guo}},\ }\href
  {\doibase 10.1038/s41534-017-0016-4} {\bibfield  {journal} {\bibinfo
  {journal} {npj Quantum Information}\ }\textbf {\bibinfo {volume} {3}},\
  \bibinfo {pages} {1} (\bibinfo {year} {2017})}\BibitemShut {NoStop}%
\bibitem [{\citenamefont {Schwemmer}\ \emph {et~al.}(2015)\citenamefont
  {Schwemmer}, \citenamefont {Knips}, \citenamefont {Richart}, \citenamefont
  {Weinfurter}, \citenamefont {Moroder}, \citenamefont {Kleinmann},\ and\
  \citenamefont {G{\"u}hne}}]{schwemmerSystematicErrorsCurrent2015}%
  \BibitemOpen
  \bibfield  {author} {\bibinfo {author} {\bibfnamefont {C.}~\bibnamefont
  {Schwemmer}}, \bibinfo {author} {\bibfnamefont {L.}~\bibnamefont {Knips}},
  \bibinfo {author} {\bibfnamefont {D.}~\bibnamefont {Richart}}, \bibinfo
  {author} {\bibfnamefont {H.}~\bibnamefont {Weinfurter}}, \bibinfo {author}
  {\bibfnamefont {T.}~\bibnamefont {Moroder}}, \bibinfo {author} {\bibfnamefont
  {M.}~\bibnamefont {Kleinmann}}, \ and\ \bibinfo {author} {\bibfnamefont
  {O.}~\bibnamefont {G{\"u}hne}},\ }\href@noop {} {\bibfield  {journal}
  {\bibinfo  {journal} {Physical Review Letters}\ ,\ \bibinfo {pages} {6}}
  (\bibinfo {year} {2015})}\BibitemShut {NoStop}%
\bibitem [{Note2()}]{Note2}%
  \BibitemOpen
  \bibinfo {note} {The validity constraint is implied and will not be noted
  henceforth.}\BibitemShut {Stop}%
\bibitem [{\citenamefont {Clauser}\ \emph {et~al.}(1969)\citenamefont
  {Clauser}, \citenamefont {Horne}, \citenamefont {Shimony},\ and\
  \citenamefont {Holt}}]{clauserProposedExperimentTest1969}%
  \BibitemOpen
  \bibfield  {author} {\bibinfo {author} {\bibfnamefont {J.~F.}\ \bibnamefont
  {Clauser}}, \bibinfo {author} {\bibfnamefont {M.~A.}\ \bibnamefont {Horne}},
  \bibinfo {author} {\bibfnamefont {A.}~\bibnamefont {Shimony}}, \ and\
  \bibinfo {author} {\bibfnamefont {R.~A.}\ \bibnamefont {Holt}},\ }\href
  {\doibase 10.1103/PhysRevLett.23.880} {\bibfield  {journal} {\bibinfo
  {journal} {Physical Review Letters}\ }\textbf {\bibinfo {volume} {23}},\
  \bibinfo {pages} {880} (\bibinfo {year} {1969})}\BibitemShut {NoStop}%
\end{thebibliography}%

\end{document}